\documentclass{an_art}
\usepackage{psfig}
\begin{document}

\begin{titlepage}
\setcounter{page}{000}
\headnote{Astron.~Nachr.~000 (0000) 0, 000--000}
\makeheadline

\title {Compact Groups of Galaxies in the Las Campanas 
Redshift Survey}

\author{{\sc S.\ S.\ Allam\footnote{Visiting Scientist, Fermi 
National Accelerator Laboratory}}, 
        Helwan, Cairo, Egypt \\
\medskip
{\small National Research Institute of Astronomy and Geophysics} \\
\bigskip
{\sc D.\ L.\ Tucker}, Batavia, Illinois, USA \\
\medskip
{\small Fermi National Accelerator Laboratory} \\
}

\date{Received ; accepted } 
\maketitle

\summary We have recently extracted a catalog of compact groups of
galaxies (CGs) from the Las Campanas Redshift Survey. This catalog of
Las Campanas Compact Groups (LCCGs) contains 76 CGs with a median
redshift of $z_{\rm med} \approx 0.08$. The physical properties of
these CGs are similar to those from the Hickson (1982) and the Barton
et al.\ (1996) catalogs. Here, we present an atlas of our catalog and
briefly describe its general properties.  END

\keyw
catalogs -- surveys -- galaxies:   galaxiesEND
\end{titlepage}

\section{Introduction}

Perhaps over half of all galaxies lie within groups containing 3
-- 20 members (Tully 1987); yet, due to the difficulty of discerning
them from the field, groups of galaxies are, as a whole, not as well
studied as larger galaxy systems.  Compact groups (CGs), however,
defined by their small number of members ($< 10$), their compactness
(typical intra-group separations of a galaxy diameter or less), and
their relative isolation (intra-group separations $\ll$ group-field
separations) are more readily identifiable; Stephan's Quintet is a
beautiful and noteworthy example.

The best known catalog of such systems is that of the Hickson
Compact Groups (HCGs; Hickson 1982), a sample comprising 100 groups
selected from the red prints of the Palomar Observatory Sky Survey
(POSS).  Due to their high densities (equivalent to those at the
centers of rich clusters) and low velocity dispersions ($\sim
200$~km~s$^{-1}$), HCGs represent an environment where interactions,
tidally triggered activity, and galaxy mergers are expected to be more
frequent than in most other surroundings.  Therefore, the HCG
catalog has served as an excellent laboratory for the study of the
effects of immediate environment on a galaxy's morphology and stellar
composition.

Nonetheless, it is necessary to compare and extend the results gleaned
from the HCG catalog using more modern catalogs. The Las Campanas
Compact Group (LCCG) catalog, which we extracted from the Las Campanas
Redshift Survey (LCRS; Shectman et al.\ 1996), contains 76 CGs and is
among the small handful of new catalogs which is extending our
knowledge of CG environments.  Since it is based upon a modern galaxy
redshift survey which in turn is based upon CCD photometry, the LCCG
catalog enjoys certain advantages over the HCG, one of the most
important being the homogeneity of the sample: e.g., one can compare
the properties of compact group galaxies {\em from the LCRS} with
those of loose group galaxies {\em from the LCRS} with those of field
galaxies {\em from the LCRS}.  Systematic errors obtained from
comparing heterogeneous samples are thus eliminated [see, for example,
the results of Allam et al.\ (1999)].

%

Here, we present both the LCCG catalog itself and a ``postage stamp''
atlas of its member groups.  We have organized the paper as follows:
in Sec.~(\ref{Data}) we overview the LCRS, in Sec.~(\ref{Selection})
we describe the ``friends-of-friends'' percolation algorithm used in
extracting the CG catalog from the LCRS, in Sec.~(\ref{Catalog}) we
present the Las Campanas Compact Group (LCCG) catalog and atlas, and
in Sec.~(\ref{Properties}) we analyze the properties of the LCCGs and
compare them to the properties of two other catalogs.  Finally, we
summarize our results in Sec.\ (\ref{Conclusions}).

\section{The Data}\label{Data}

The LCRS is an $R$-band selected survey; it extends to a redshift of
0.2 and contains 26,418 galaxy redshifts of which 23,697 lie within
the official geometric and photometric limits of the survey.  Accurate
photometry and sky positions for program objects were extracted from
CCD drift scans obtained on the Las Campanas Swope 1-m telescope;
spectroscopy was performed at the Las Campanas Du~Pont 2.5-m
telescope, originally via a 50-fiber Multi-Object Spectrograph (MOS),
and later via a 112-fiber MOS.

For observing efficiency, all the fibers were used, but each MOS field
was observed only once.  Hence, the LCRS is a collection of 50-fiber
fields (with nominal apparent magnitude limits of $16.0 \leq R <
17.3$) and 112-fiber fields (with nominal apparent magnitude limits of
$15.0 \leq R < 17.7$).  Observing each field only once, however,
creates a problem for finding compact groups: the protective tubing of
the individual fibers prevented the spectroscopic observation of both
members of galaxy pairs within 55~arcsec of each other.  Therefore,
many galaxies in CG environments are missing from the LCRS redshift
catalog.  We have partially circumvented this problem by assigning
each of the $\sim$ 1,000 ``missing'' LCRS galaxies the redshift of its
nearest neighbor (convolved with a gaussian of
$\sigma$=200~km~s$^{-1}$). Hence, on the small angular scales
necessary for compact group selection, the LCRS falls somewhere
between a 2D sky survey and a fully 3D redshift survey.

\section{Compact Group Selection}\label{Selection}

We have selected CGs based on physical extent rather than angular
size.  We applied a friends-of-friends algorithm (Ramella et al.\
1989) to extract a sample of CGs systems in the LCRS (hereafter,
LCCGs).  In constructing the LCCG catalog, we have considered only
those LCRS galaxies within the official geometric and photometric
borders of the survey; furthermore, we have limited this sample to
galaxies having redshifts in the range $7,500~\mbox{km~s}^{-1} \leq
cz_{\rm cmb} < 50,000~\mbox{km~s}^{-1}$ and luminosities in the range
$-22.5\leq M_R - 5\log h < -17.5$. To avoid CG member incompleteness
at the extremal distances of the sample, only groups within the range
$10,000~\mbox{km~s}^{-1} \leq cz_{\rm cmb} < 45,000~\mbox{km~s}^{-1}$
were admitted into the final LCCG catalog.

We identify CGs as linked sets of neighboring galaxies.  A seed galaxy
$i$ is selected which has not yet been classified as either a CG
member or an isolated galaxy.  Every other non-classified galaxy
$j$ in the survey sample is then tested to see if it lies within a
projected separation $D_{\rm L}$ and a velocity difference $V_{\rm L}$
of the seed galaxy:
\vspace{-.18cm}
\begin{equation}\label{eq1}
D_{ij} = 2 D_{\rm ave} \sin(\Theta_{ij}/2) \leq D_{\rm L} \mbox{ and }
\vspace{-.18cm}
\end{equation}
\begin{equation}\label{eq2}
V_{ij} = c \times | z_i - z_j | \leq V_{\rm L} , 
\end{equation}
\vspace{-.18cm} 
where $D_{\rm ave} \equiv ( D(z_i) + D(z_j) ) / 2$ and $\Theta_{ij}$
is the angular separation between the two galaxies.  The comoving
distances $D(z)$ are calculated assuming $q_0 = 0.5$ and $H_0 = h
\times 100$~km~s$^{-1}$~Mpc$^{-1}$ ($h \equiv
H_0/100$~km~s$^{-1}$~Mpc$^{-1}$), and the redshifts $z$ are corrected
for motion relative to the dipole moment of the cosmic microwave
background (CMB; Lineweaver et al.\ 1996).

If no companions are found within $D_{\rm L}$ and $V_{\rm L}$ of the
seed galaxy, it is assigned ``isolated'' status and another seed
galaxy is sought.  If companions are found, they are added along with
the seed galaxy to a list of CG members forming a new group.  In turn,
the surroundings of each of these companions are combed for the next
level of ``friends.''  This loop is repeated until no further
companions are located, and the process is begun again by pursuing
another seed galaxy.  The CG catalog is complete only once every
galaxy in the redshift sample has been classified as either
``isolated'' or ``grouped.''  Only those groups containing three or
more members are included in the final catalog.

Following Barton et al.\ (1996), we chose values of $D_{\rm L}$ and
$V_{\rm L}$ which would produce a catalog of LCRS compact groups which
had physical properties similar to those of the HCGs.  Our definition
was as follows:
\begin{itemize}
\vspace{-.18cm} 
\item low membership: number of group members $N_{\rm tot} \geq 3$ galaxies, 
\vspace{-.23cm} 
\item compact: projected inter--galaxy separation of $D_{ij} \le
D_{\rm L} = 50h^{-1}$~kpc ($\sim$ 1 galaxy diameter), and
\vspace{-.23cm}
\item isolated: inter--galaxy velocity difference $V_{ij} \le V_{\rm
L} = 1000$~km~s$^{-1}$
\vspace{-.18cm} 
\end{itemize}

Using these criteria a total of 76 CGs were identified by a systematic
search of the LCRS, each CG having three or more member galaxies. The
total number of galaxies contained witin the set of LCCGs is 253.

\section{Compact Group Catalog}\label{Catalog}

We list the properties of all the LCCGs in Table~(\ref{tab:Cgroup}).
Tabulated are the following:

\begin{description}

\item[Column (1):] a running identification number.
\vspace{-0.18cm}
\item[Columns (2 \& 3):] the group's right ascension and declination 
                        (1950 coordinates).
\vspace{-0.18cm}
\item[Column (4):] $z_{\rm cmb}$, the group's redshift.
\vspace{-0.18cm}
\item[Column (5):] $N_{\rm tot}$, the number of galaxies in the group.
\vspace{-0.18cm}
\item[Column (6):] $R_{\rm p}$, the group's mean pairwise separation.
\vspace{-0.18cm}
\item[Column (7):] $R_{\rm h}$, the group's harmonic radius.
\vspace{-0.18cm}
\item[Column (8):] $L_{\rm tot}$, the group's LCRS $R$-band total luminosity, 
                   corrected for galaxies outside the LCRS's photometric limits.
                   (Since several other group catalogs have photometry in
                   the de~Vaucouleurs $B_{0}$ band, it is useful to note
                   this rough conversion relation:  
                   $L_{\rm LCRS} \approx 1.1 L_{B_{0}}$.)  
         
\vspace{-0.18cm}
\item[Column (9):] $L_{\rm rat}$, the ratio by which the observed group members' 
                   summed luminosities must be multiplied to obtain $L_{\rm tot}$.
\vspace{-0.18cm}
\item[Column (10):] $\Theta_{\rm G}$, the angular radius of the smallest circle which 
                    includes all the group's members.
\vspace{-0.18cm}
\item[Column (11):] $R_{\rm G}$, the physical radius of the smallest circle which 
                    includes all the group's members.
\vspace{-0.18cm}
\item[Column (11):] the type of field in which the group sits 
                   (``1'' $=$ a 50-fiber field, ``2'' $=$ a 112-fiber field).
\vspace{-0.18cm}
\item[Column (12):]  A column to reference any applicable notes for the
                     group:
                     \vspace{-0.18cm}
                     \begin{description}
                     \item[a:] the group's barycenter is closer than $2R_{\rm p}$ 
                               to a slice edge.
                     \vspace{-0.18cm}
                     \item[b:] the group contains at least one 55~arcsec ``orphan'' 
                               with a mock redshift.
                     \end{description}
\end{description}

Further details in how these properties were measured can be found in
Tucker et al.\ (2000), which describes a catalog of LCRS {\em loose\/}
groups.

\subsection{Group Members}

Table~\ref{tab:Cgmem} lists the members of each group. Column (1)
provides a group member identification, where the letter ``a''
represents the brightest group member, the letter ``b'' the second
brightest group member, and so on.  Columns (2~\&~3) list the group
member's right ascension and declination in 1950.0 coordinates. The
LCRS $R$-band isophotal magnitude for the group member is given in
Column (4), and its velocity is listed in Col.\ (5).  As noted
earlier, in generating the LCLG catalog, each 55-arcsec ``orphan'' was
assigned a mock velocity equal to the gaussian-convolved velocity of
its nearest neighbor (a gaussian of $\sigma$=200~km~s$^{-1}$ was
used); this mock velocity is listed in this column for 55-arcsec
``orphan'' group members.

\begin{figure}[ht]
\centering
\makebox[60mm]{\psfig{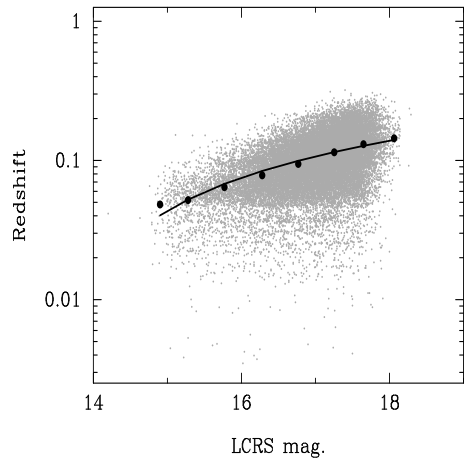}}
\caption{$R$-band magnitude vs.\ measured (i.e., spectroscopic)
redshift for galaxies in the LCRS spectroscopic sample.  The solid
line represents the linear fit. The filled circles denote the median
redshift for a given magnitude bin.}
\label{fig:magz}
\end{figure}

Using the $R$-band magnitude of the LCRS ($m_{R}$) and the measured
spectroscopic redshift, a photometric redshift relation was also
derived. A first order least square fit was made to the data
(0.03$\pm$0.001 $\times m_{R}$ - 0.43$\pm$0.028;
see Fig.\ \ref{fig:magz}).  This magnitude--redshift relation was then
re--applied to each CG member with a mock velocity; the resulting
photometric redshift, $z_{\rm photo}$, is listed in Col.\ (6).

Finally, in Col.\ (7), we provide a qualitative description of the
group member's spectrum:  
\begin{description}
\item[c:] a ``continuum'' spectrum (i.e., a spectrum with no strong
          emission line features),
\vspace{-0.18cm}

\item[e:] a spectrum containing strong emission lines,
\vspace{-0.18cm}
\item[b:] a spectrum which shows both absorption lines (as in a continuum
          spectrum) and weak emission lines, and 
\vspace{-0.18cm}
\item[m:] a 55-arcsec ``orphan'' with a mock redshift.
\end{description}

\subsection{The LCCG Atlas} 

For each of the 76 LCCGs, we have created an atlas image with the
following characteristics (see Fig.\ \ref{fig:cgs}):
\vspace{-.18cm}
\begin{itemize}
\item Each atlas image is a 100~pixel$\times$100~pixel ``postage
      stamp'' with a resolution of 1.7~arcsec~pixel$^{-1}$ (obtained
      from the Digitized Sky Survey, courtesy of SkyView).
\vspace{-.23cm}
\item Within each atlas image, the identified group members are
      labelled by their lowercase letter designation (a, b, c, ...; 
      see Table~\ref{tab:Cgmem}).
\vspace{-.23cm}
\item Within each atlas image, the group barycenter is marked by a
      ``$\times$''.
\end{itemize}
\vspace{-.18cm}
Note that in many of these fields there are galaxies which are {\em
not\/} identified as group members.  These galaxies are either 
outside the photometric limits of the LCRS redshift catalog, and/or
members of a loose group or rich cluster in which the LCCG is
embedded, and/or 55~arcsec ``orphans'' which have been accidentally
excluded from the LCCG.

\section{Properties of LCCGs}\label{Properties}

\subsection{Compact Groups and Large--Scale Structure}

\begin{figure}[ht] \begin{center}\begin{picture}(300,300)
\put(-65,-80){\includegraphics{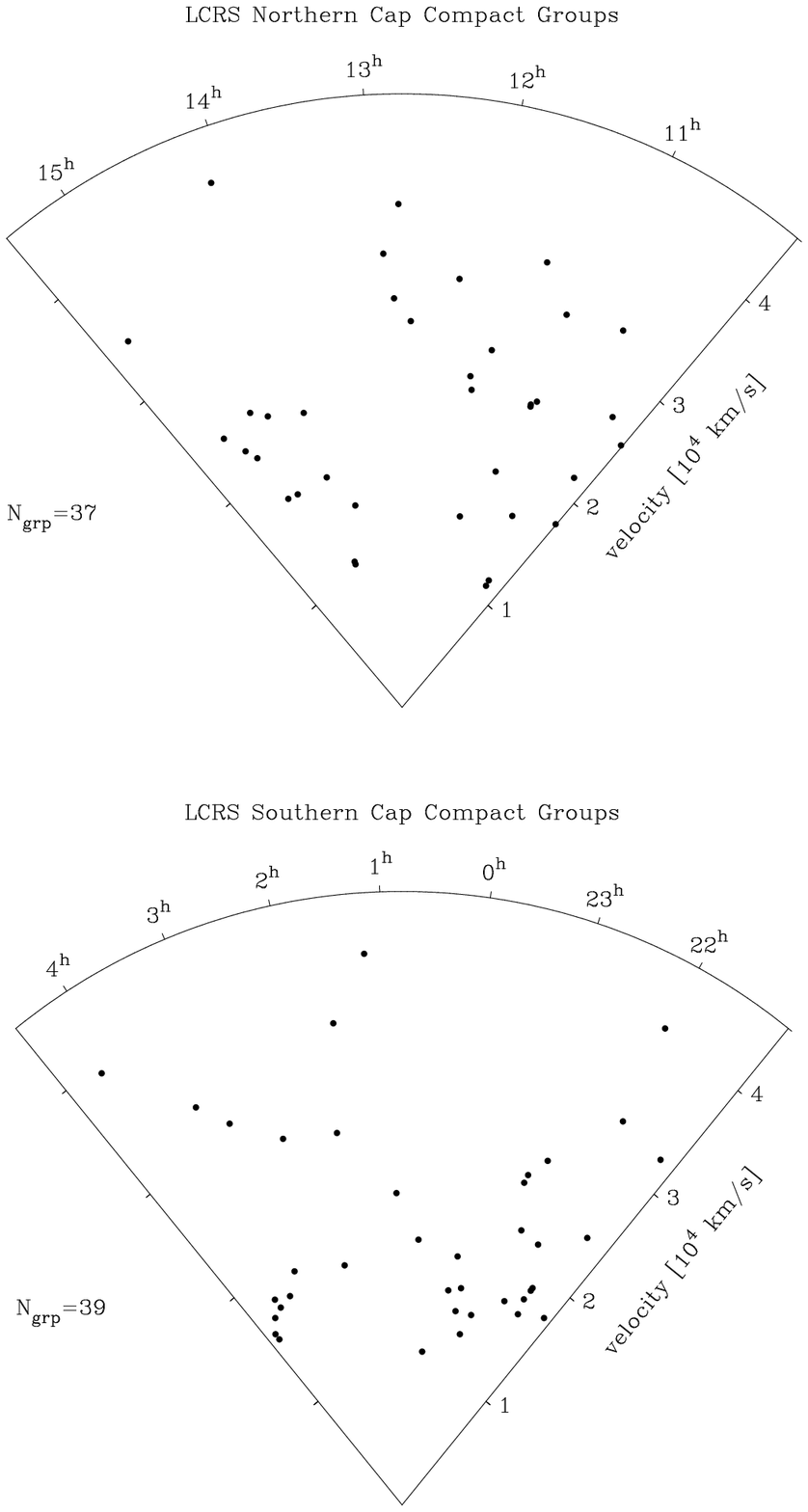}}
\put(120,-80){\includegraphics{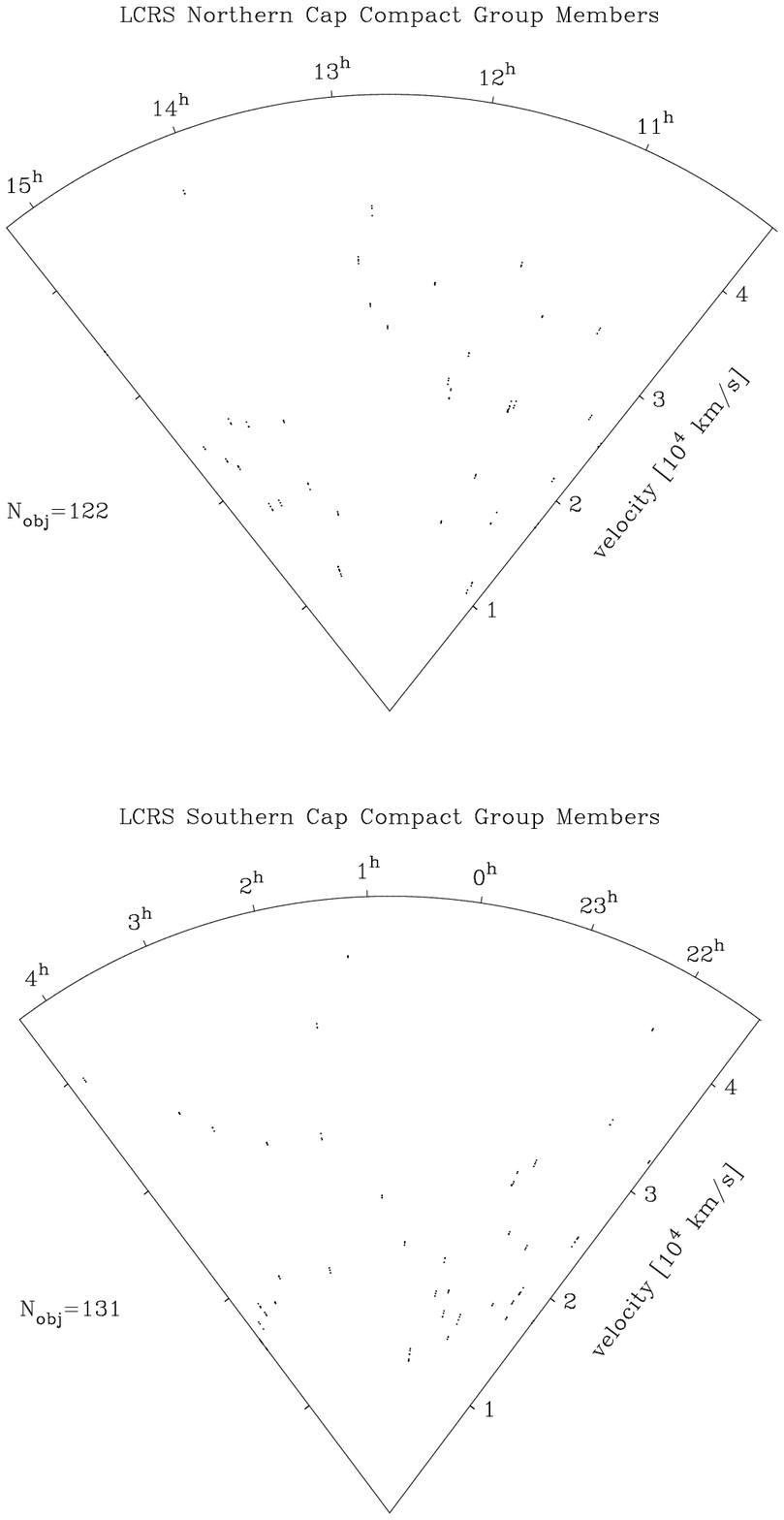}}
\put(20,150){\large{\bf A.}}\put(210,150){\large{\bf B.}}
\put(20,-10){\large{\bf C.}}\put(210,-10){\large{\bf D.}}
\end{picture}\end{center}
\caption{The cone diagrams of LCCGs in the Northern (top), and 
Southern (bottom) Galactic Cap.}
\label{fig:gr}\end{figure}

The cone diagrams in Figures (\ref{fig:gr}A--D) show the positions of
the LCCGs and their individual group members for the LCRS Northern and
Southern Galactic Cap regions.  Note that the distribution of LCCGs is
non-random: from comparison with cone diagrams of all LCRS galaxies
(Shectman et al.\ 1996), it is clear that the LCCGs (and the LCCG
members) trace many of the same large-scale structures, albeit much
more sparsely.  This is not too surprising, since one would not expect
to find CGs where where there are no galaxies.  On the other hand,
this adds fuel to the arguments that CGs are not truly isolated, and
that many of them may just be chance alignments along the
line-of-sight of large-scale filamentary structures and/or that they
may form continuously within loose groups or rich clusters of galaxies
(Mamon 1986, Hernquist et al.\ 1995, Diaferio et al.\ 1994).  [Due to
the significant differences between the star formation properties of
loose group galaxies and compact group galaxies (e.g., Allam et al.\
1999), the latter scenario -- in which CGs are real substructures
within loose groups and rich clusters and not merely chance alignments
-- is the more likely.]

\subsection{Environments of Compact Groups}

We have examined the regions around the LCCGs visually using images
from the Digitized Sky Survey and by using the NED database.  We have
also correlated the positions of these CGs with the catalog of loose
groups of galaxies by Tucker et al.\ (2000).  From these
investigations, we have found the following:
\vspace{-.18cm}
\begin{itemize}
\item $\sim 1/3$ of LCCGs are ``isolated,'' lying outside group or cluster
environments,
\vspace{-.23cm}
\item $\sim 1/3$ of LCCGs are embedded within poor loose groups of galaxies,
and 
\vspace{-.23cm}
\item $\sim 1/3$ of LCCGs are embedded within rich, Abell-class clusters.
\end{itemize}
\vspace{-.18cm} 


\subsection{Comparison with Other Compact Group Catalogs}

\begin{figure}[ht]
\centering
\makebox[55mm]{\psfig{file=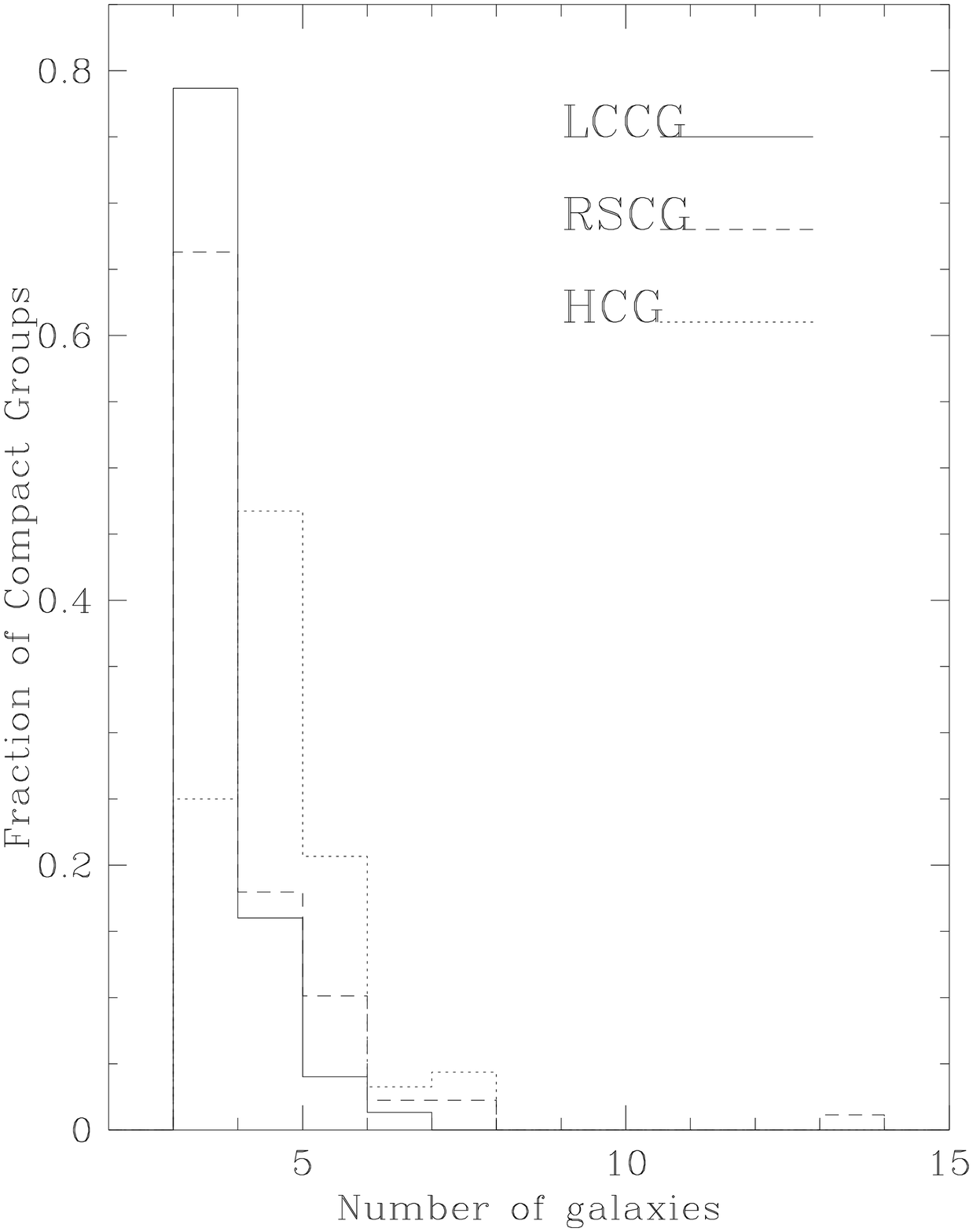,width=55mm,height=55mm,silent=}{\large{\bf
a}}}
\makebox[55mm]{\psfig{file=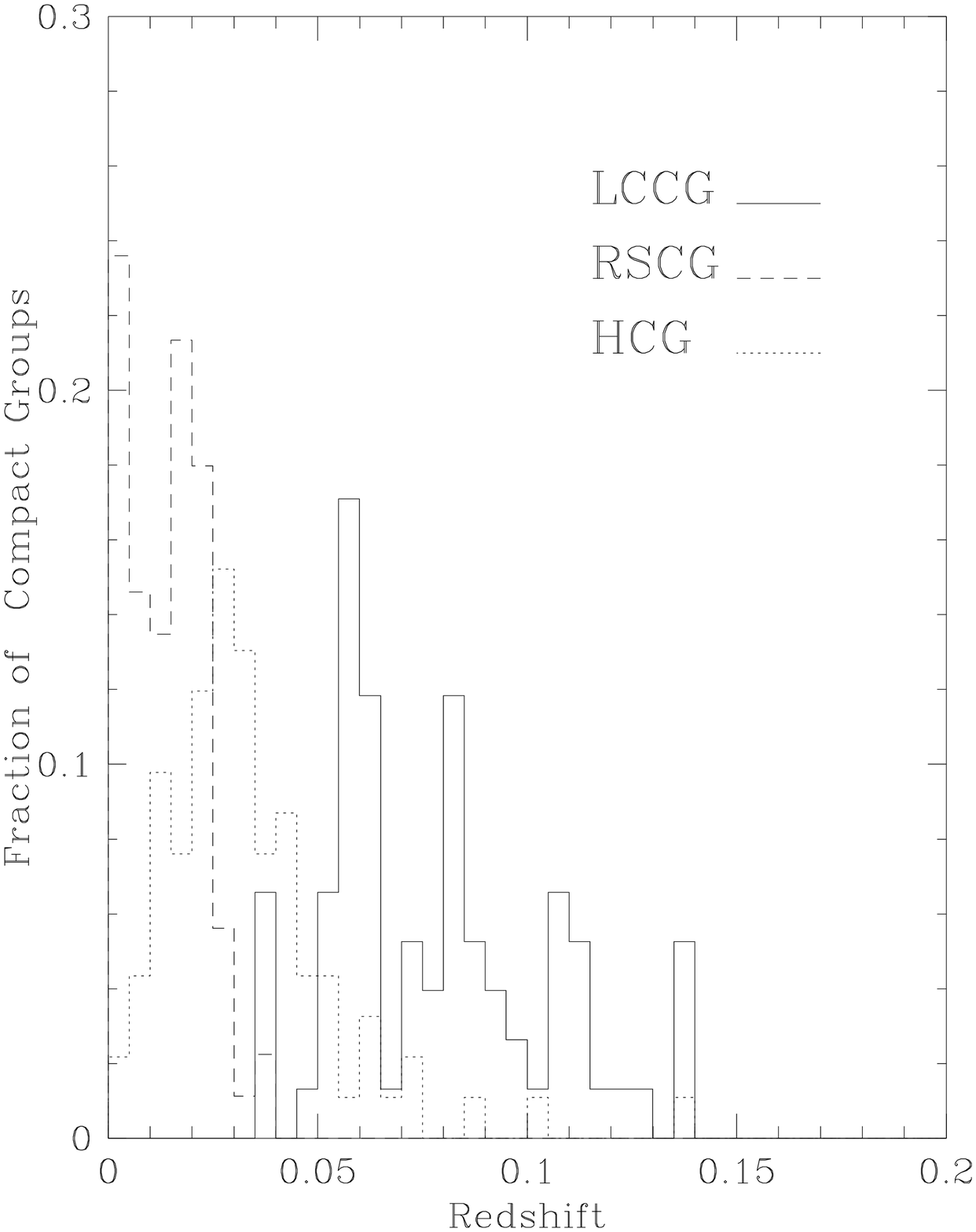,width=55mm,height=55mm,silent=}{\large{\bf
b}}}
\makebox[55mm]{\psfig{file=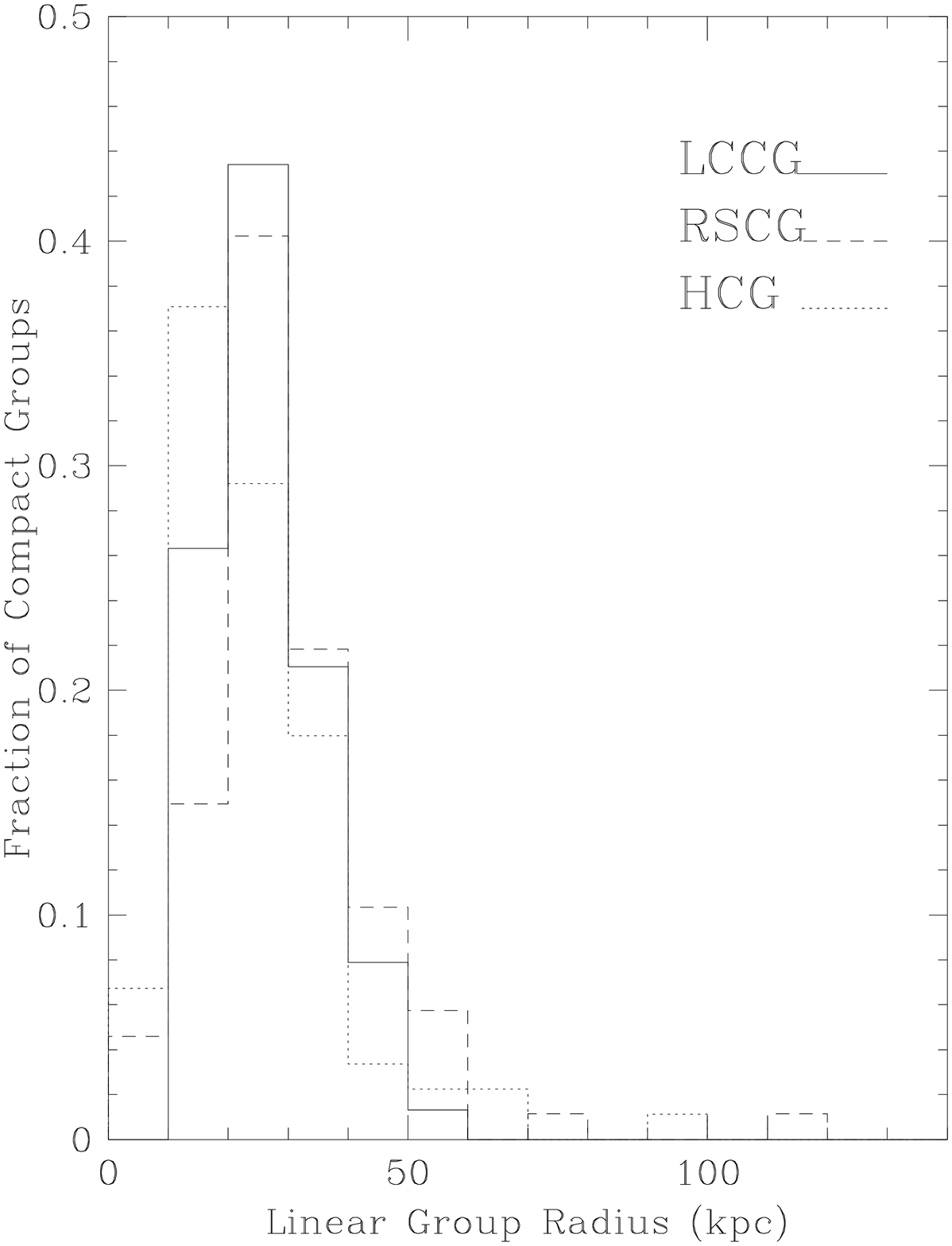,width=55mm,height=55mm,silent=}{\large{\bf
c}}}
\makebox[55mm]{\psfig{file=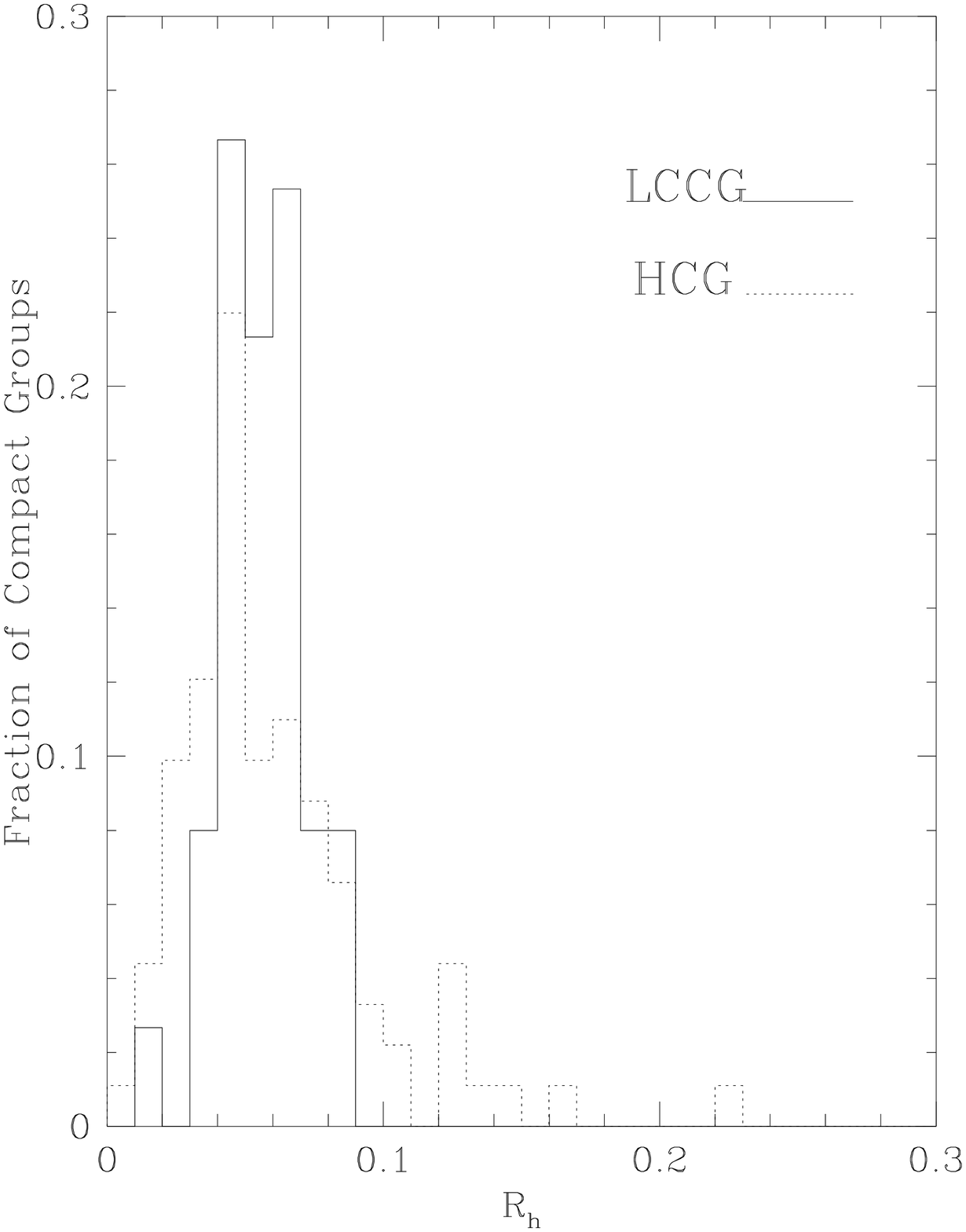,width=55mm,height=55mm,silent=}{\large{\bf
d}}}
\makebox[55mm]{\psfig{file=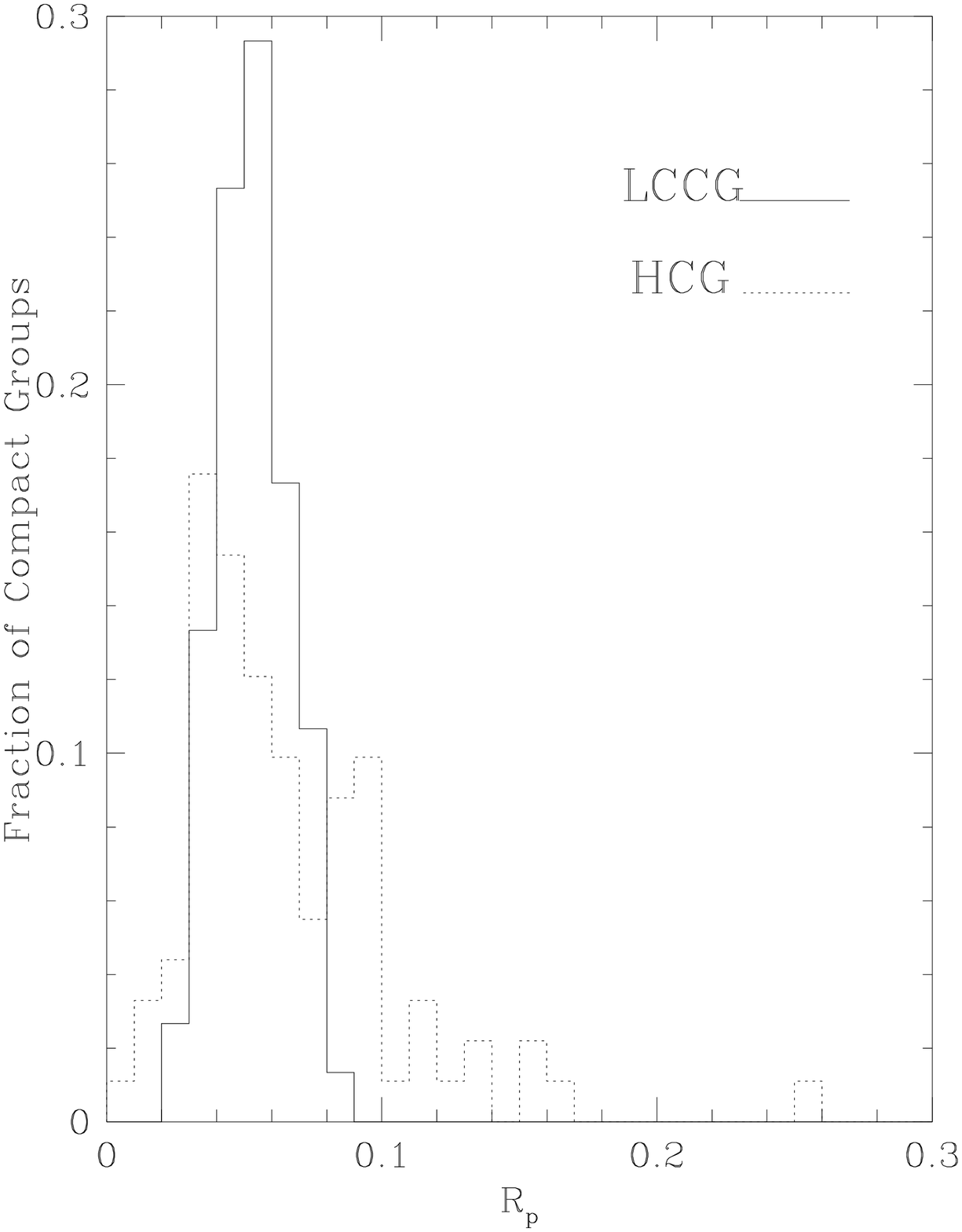,width=55mm,height=55mm,silent=}{\large{\bf
e}}}
\makebox[55mm]{\psfig{file=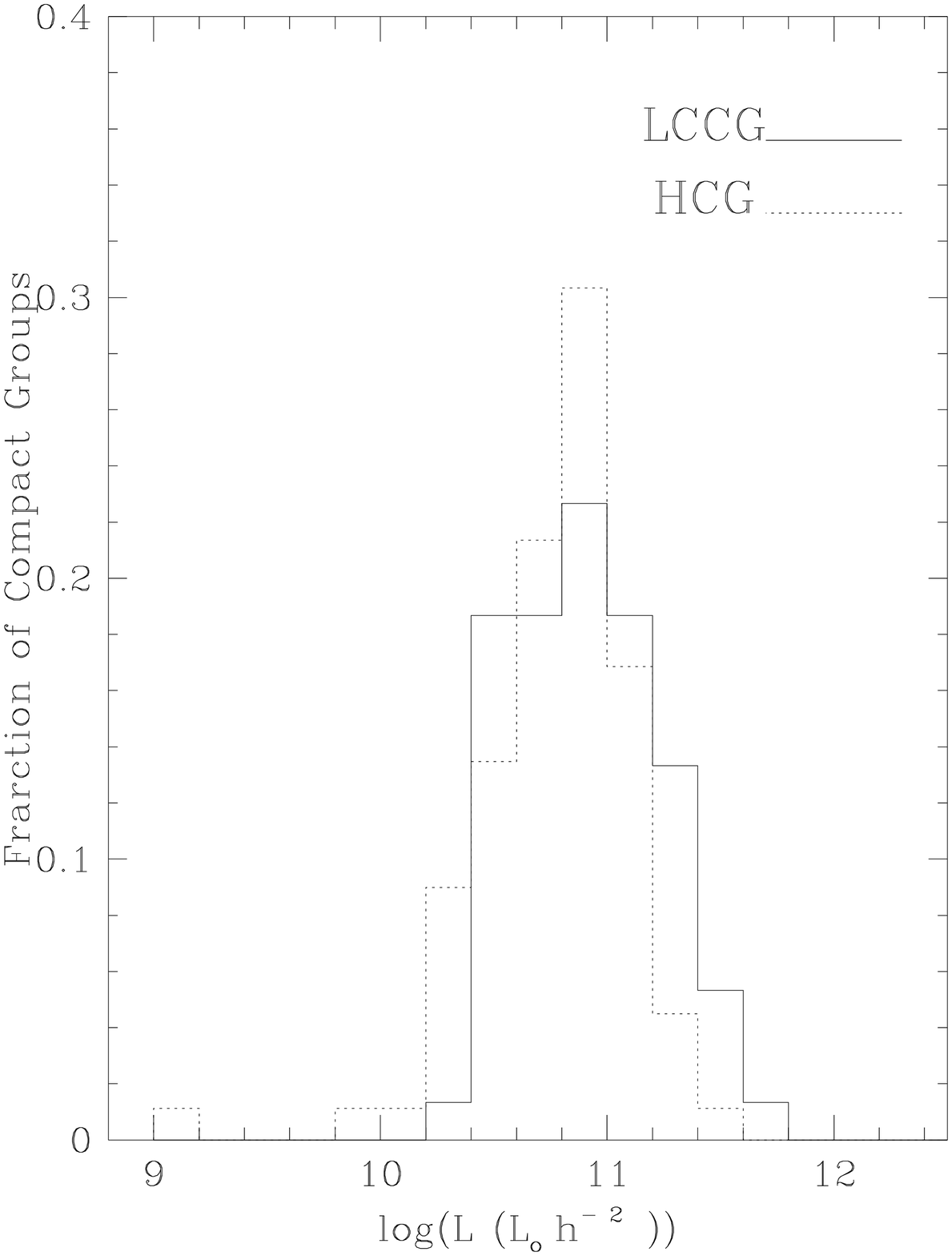,width=55mm,height=55mm,silent=}{\large{\bf
f}}}
\caption{(a) The distribution of group populations, $N_{\rm tot}$, (b)
the distribution of group redshifts, $z_{\rm cmb}$, (c) the distribution
of linear group radii, $R_{\rm G}$, (d) the distribution of harmonic
radii, $R_{\rm h}$, (e) the distribution of mean pairwise separations,
$R_{\rm p}$, and (f) the distribution of total group luminosities, $L_{\rm
tot}$, for LCCGs (solid line), RSCGs (dashed line), and HCGs (dotted
line).  Each distribution is normalized to the total number of CGs in
each sample.  For (f), the luminosities of the HCGs were converted to
LCRS $R$-band luminosities assuming $L_{\rm LCRS} \approx 1.1 L_{B_{0}}$.}
\label{fig:stat}
\end{figure}

It is instructive to compare the LCCG catalog with CG catalogs derived
from other galaxy redshift surveys, like that by Barton et al.\ (1996)
(the RSCG catalog), and those based upon sky surveys (e.g., the HCG
catalog).  Table~\ref{tab:Cgstat} and Figure~\ref{fig:stat} summarize
the properties of the 76 LCCGs, the 92 HCGs, and the 89 RSCGs.  Note
that, for the most part, all three catalogs contain systems with very
similar physical properties.  There are, however, two areas in which
the catalogs differ.  First, the HCG catalog contains a smaller
fraction of triplets than do either the LCCG or the RSCG catalogs, a
result of a different population threshold originally for the HCG
catalog.  (Groups were only admitted to the HCG catalog from the POSS
plates if they contained at least four galaxies; groups were only
admitted to the LCLG and the RSCG samples if they contained at least
three galaxies.)  Second, the LCCG catalog is much deeper on average
than either of the other two CG catalogs, which is due to the depth of
the LCRS itself.

\begin{table}[hp]
\caption[]{Compact Groups Physical Parameters}
\label{tab:Cgstat}
\begin{flushleft}
\begin{center} {\small
\begin{tabular}{lcccccc} 
\hline
\noalign{\smallskip}
Parameter &\multicolumn{2}{c}{LCCGs}
&\multicolumn{2}{c}{HCGs}& \multicolumn{2}{c}{RSCGs}\\
&  mean &median& mean& median& mean &median\\
\noalign{\smallskip}
\hline
\noalign{\smallskip}		
$N_{\rm tot}$                   & 3.28$\pm$0.069  & 3     & 4.15$\pm$0.10    & 4     & 3.65$\pm$0.14  & 3 \\
$z_{\rm cmb}$                   & 0.08$\pm$0.003  & 0.079 & 0.03$\pm$0.002   & 0.03  & 0.013$\pm$0.001& 0.014 \\
$\theta_{\rm G}$ [deg]          & 0.016$\pm$0.0009& 0.0131& 0.0601$\pm$0.0005& 0.0483&  ...           & ...  \\ 
$R_{\rm G}$ [$h^{-1}$~kpc]& 26.31$\pm$1.00  & 26.0  & 24.69$\pm$1.53   & 21.76 & 29.99$\pm$1.63 & 26.90 \\ 
$R_{\rm p}$      [$h^{-1}$~Mpc] & 0.05$\pm$0.002  & 0.05  & 0.05$\pm$0.02    & 0.05  &  ...           & ... \\
$R_{\rm h}$      [$h^{-1}$~Mpc] & 0.05$\pm$0.002  & 0.05  & 0.06$\pm$0.004   & 0.05  &  ...            & ... \\
$L$ [$10^{10}L_{\odot}h^{-2}$]  & 7.24$\pm$1.08   & 7.4$^*$   & 7.4$\pm$0.54$^*$     & 6.5   &  ...            & ... \\
\noalign{\smallskip}
\hline
\multicolumn{7}{l}{$^*$ The HCG total luminosities were converted to the LCRS $R$-band assuming $L_{\rm LCRS} \approx 1.1 L_{B_{0}}$.}
\end{tabular} }
\end{center}
\end{flushleft}
\end{table}


\section{Conclusions}\label{Conclusions}

We have applied a ``friends-of-friends'' group identification
algorithm (Huchra \& Geller 1982) to the Las Campanas Redshift Survey
(LCRS) in order to produce a catalog of compact groups (CGs)
physically similar to those of the Hickson (1982) catalog. We defined
these Las Campanas Compact Groups (LCCGs) by the following criteria:
$N_{\rm tot} \geq 3$ galaxies, with projected separations of $D \leq
D_{\rm L} = 50h^{-1}$~kpc ($\sim$ 1 galaxy diameter) and velocity
differences of $V \leq V_{\rm L} = 1000$~km~s$^{-1}$.  As a result, we
extracted a catalog of 76 LCCGs, each containing 3 or more galaxies.
The total number of galaxies in the LCCG sample is 253.  All the LCCGs
contain at least 1 spectroscopically determined redshift; 23 contain 2
or more. The mean redshift of LCCGs is $z_{\rm cmb} \sim 0.08$.


We have evaluated the physical characteristics of the LCCGs and have
found the following:
\begin{itemize}
\vspace{-.25cm} 
\item The main physical properties of the LCCGs (membership, linear
size, total group luminosity) are similar to those of HCGs.
\vspace{-.25cm} 
\item The median redshift is $\sim$0.08, more than
twice that of either the HCG or the Barton et al.\ (1996) catalogs.
\vspace{-.25cm} 
\item Only about one-third of LCCGs are isolated; the remaining
two-thirds are embedded within loose groups or clusters.
\vspace{-.25cm} 
\end{itemize}

We therefore conclude that the catalog of LCCGs is a useful addition
to the study of compact groups.  We further conclude that, although a
significant minority of compact groups appear to be truly isolated,
the majority of these systems are in reality dense substructures
within larger galaxy associations (e.g., loose groups and rich
clusters).

\begin{table*}[ht]
\begin{center}
 \caption[c]{The main properties of the Las Campanas Compact Groups (LCCGs).}
 \label{tab:Cgroup}
 \begin{flushleft}
 \begin{center} {\small
\begin{tabular}{lc@{\,}c@{\,}c@{ }c@{\,}c@{\,} ccccccccccc}
\hline
\noalign{\smallskip}
N & \multicolumn{3}{c}{RA} & \multicolumn{3}{c}{DEC} & $z_{cmb}$  & $N_{tot}$ & $R_{P}$ &$R_{h}$ & L & $L_{rat}$ & $\theta_{G}$ & $R_{G}$& Type & Notes \\
 \noalign{\smallskip}
&\multicolumn{6}{c}{(1950)}& & &\multicolumn{2}{c}{Mpc~$h^{-1}$} & $L_{\odot}~h^{-2}$ & & &Mpc~$h^{-1}$ &&\\
 \noalign{\smallskip}
(1) & \multicolumn{3}{c}{(2)} & \multicolumn{3}{c}{(3)} & (4) & (5) & (6) & (7) & (8) & (9) & (10) & (11) & (l2) & (13)  \\
 \noalign{\smallskip}
\hline
 \noalign{\smallskip}
01 & 00 & 22 & 04.59 & -45 & 34 & 16.0 & 0.0666 & 3 & 0.04 & 0.05 & 0.63E11 & 1.91 & 0.0132 & 0.0411 & 2 & b  \\
02 & 00 & 45 & 49.54 & -42 & 17 & 07.0 & 0.0780 & 4 & 0.07 & 0.08 & 0.18E12 & 6.12 & 0.0245 & 0.0875 & 1 & b  \\
03 & 01 & 04 & 07.90 & -44 & 35 & 11.9 & 0.1384 & 3 & 0.04 & 0.05 & 0.15E12 & 3.01 & 0.0064 & 0.0370 & 2 & b  \\
04 & 01 & 21 & 20.48 & -39 & 29 & 51.7 & 0.1217 & 3 & 0.06 & 0.06 & 0.18E12 & 4.64 & 0.0112 & 0.0585 & 2 & b  \\
05 & 01 & 39 & 07.37 & -44 & 51 & 02.5 & 0.0948 & 3 & 0.05 & 0.05 & 0.55E11 & 2.17 & 0.0121 & 0.0511 & 2 & b  \\
06 & 02 & 00 & 24.78 & -45 & 01 & 18.8 & 0.0620 & 3 & 0.04 & 0.04 & 0.29E11 & 1.98 & 0.0114 & 0.0332 & 2 & b  \\
07 & 02 & 27 & 01.50 & -44 & 40 & 42.5 & 0.0972 & 3 & 0.06 & 0.08 & 0.16E12 & 4.34 & 0.0142 & 0.0615 & 1 & b  \\
08 & 02 & 50 & 11.91 & -39 & 06 & 40.7 & 0.1050 & 3 & 0.04 & 0.05 & 0.12E12 & 2.46 & 0.0095 & 0.0436 & 2 & b  \\
09 & 02 & 53 & 17.43 & -39 & 26 & 33.2 & 0.0646 & 3 & 0.05 & 0.06 & 0.71E11 & 1.79 & 0.0185 & 0.0561 & 2 & b  \\
10 & 03 & 18 & 32.39 & -42 & 26 & 33.9 & 0.1138 & 3 & 0.06 & 0.07 & 0.16E12 & 5.02 & 0.0131 & 0.0646 & 2 & b  \\
11 & 03 & 20 & 27.75 & -41 & 32 & 01.7 & 0.0601 & 4 & 0.05 & 0.04 & 0.38E11 & 4.31 & 0.0261 & 0.0741 & 2 & b  \\
12 & 03 & 30 & 26.56 & -38 & 18 & 06.3 & 0.0585 & 3 & 0.02 & 0.03 & 0.63E11 & 2.72 & 0.0092 & 0.0256 & 2 & b  \\
13 & 03 & 33 & 22.62 & -38 & 50 & 10.9 & 0.0611 & 4 & 0.07 & 0.08 & 0.89E11 & 3.43 & 0.0294 & 0.0846 & 2 & b  \\
14 & 03 & 57 & 48.08 & -41 & 28 & 00.8 & 0.0581 & 3 & 0.07 & 0.08 & 0.67E11 & 1.72 & 0.0293 & 0.0808 & 2 & ab  \\
15 & 04 & 17 & 01.69 & -42 & 17 & 12.5 & 0.0554 & 3 & 0.05 & 0.05 & 0.28E11 & 1.86 & 0.0197 & 0.0520 & 2 & b  \\
16 & 04 & 17 & 08.65 & -42 & 19 & 16.7 & 0.0537 & 3 & 0.07 & 0.07 & 0.60E11 & 1.92 & 0.0281 & 0.0720 & 2 & b  \\
17 & 04 & 19 & 34.59 & -45 & 22 & 10.0 & 0.1391 & 3 & 0.03 & 0.04 & 0.24E12 & 4.67 & 0.0049 & 0.0283 & 2 & b  \\
18 & 10 & 04 & 30.11 & -05 & 22 & 43.2 & 0.0598 & 3 & 0.03 & 0.03 & 0.67E11 & 2.37 & 0.0085 & 0.0241 & 2 & b  \\
19 & 10 & 04 & 52.96 & -05 & 26 & 35.4 & 0.0854 & 3 & 0.04 & 0.05 & 0.12E12 & 2.53 & 0.0120 & 0.0465 & 2 & b  \\
20 & 10 & 17 & 28.72 & -03 & 37 & 33.5 & 0.0718 & 3 & 0.04 & 0.04 & 0.57E11 & 2.33 & 0.0109 & 0.0364 & 2 & b  \\
21 & 10 & 20 & 42.80 & -05 & 37 & 23.6 & 0.0897 & 3 & 0.04 & 0.04 & 0.25E12 & 7.17 & 0.0094 & 0.0380 & 1 & b  \\
22 & 10 & 26 & 21.32 & -02 & 59 & 02.0 & 0.0370 & 3 & 0.05 & 0.04 & 0.23E11 & 3.41 & 0.0306 & 0.0555 & 2 & b  \\
23 & 10 & 27 & 38.03 & -02 & 54 & 40.0 & 0.0385 & 3 & 0.06 & 0.06 & 0.34E11 & 3.42 & 0.0296 & 0.0557 & 2 &   \\
24 & 10 & 43 & 30.75 & -02 & 18 & 43.3 & 0.1095 & 3 & 0.05 & 0.06 & 0.86E11 & 2.78 & 0.0105 & 0.0502 & 2 & b  \\
25 & 10 & 44 & 05.36 & -11 & 22 & 00.7 & 0.0555 & 3 & 0.04 & 0.04 & 0.30E11 & 2.01 & 0.0158 & 0.0418 & 2 & b  \\
26 & 11 & 09 & 48.15 & -02 & 53 & 10.1 & 0.0837 & 3 & 0.04 & 0.04 & 0.11E12 & 2.56 & 0.0105 & 0.0399 & 2 & b  \\
27 & 11 & 12 & 14.34 & -06 & 05 & 40.1 & 0.0818 & 4 & 0.05 & 0.06 & 0.32E12 & 6.10 & 0.0158 & 0.0588 & 1 & b  \\
28 & 11 & 12 & 49.75 & -03 & 38 & 04.0 & 0.0824 & 5 & 0.08 & 0.07 & 0.91E11 & 2.62 & 0.0299 & 0.1123 & 2 & b  \\
29 & 11 & 13 & 19.86 & -11 & 17 & 30.0 & 0.1069 & 3 & 0.05 & 0.05 & 0.13E12 & 2.85 & 0.0124 & 0.0581 & 2 & ab  \\
30 & 11 & 18 & 31.24 & -02 & 28 & 30.8 & 0.0636 & 3 & 0.03 & 0.03 & 0.29E11 & 1.52 & 0.0088 & 0.0263 & 2 & b  \\
31 & 11 & 32 & 18.84 & -12 & 21 & 49.7 & 0.1177 & 3 & 0.07 & 0.08 & 0.20E12 & 3.81 & 0.0142 & 0.0718 & 2 & b  \\
32 & 11 & 37 & 36.45 & -03 & 04 & 56.0 & 0.0500 & 3 & 0.03 & 0.01 & 0.32E11 & 2.73 & 0.0120 & 0.0289 & 2 & b  \\
33 & 11 & 48 & 38.04 & -02 & 45 & 01.6 & 0.0923 & 3 & 0.04 & 0.04 & 0.14E12 & 3.49 & 0.0089 & 0.0367 & 2 & b  \\
34 & 11 & 55 & 36.56 & -02 & 16 & 40.4 & 0.0815 & 5 & 0.07 & 0.06 & 0.91E11 & 1.58 & 0.0226 & 0.0838 & 2 & ab  \\
35 & 11 & 58 & 19.59 & -11 & 24 & 12.0 & 0.0849 & 3 & 0.05 & 0.06 & 0.41E11 & 1.79 & 0.0136 & 0.0523 & 2 & b  \\
36 & 12 & 14 & 29.53 & -02 & 55 & 20.5 & 0.1083 & 5 & 0.06 & 0.06 & 0.19E12 & 3.11 & 0.0165 & 0.0781 & 2 & b  \\
37 & 12 & 40 & 35.33 & -11 & 57 & 26.3 & 0.0971 & 3 & 0.03 & 0.04 & 0.11E12 & 2.86 & 0.0067 & 0.0288 & 2 & b  \\
38 & 12 & 47 & 00.38 & -06 & 04 & 39.3 & 0.1259 & 3 & 0.06 & 0.06 & 0.51E12 & 8.04 & 0.0124 & 0.0661 & 1 & b  \\
39 & 12 & 50 & 23.74 & -11 & 48 & 44.0 & 0.1028 & 3 & 0.04 & 0.05 & 0.66E11 & 2.72 & 0.0098 & 0.0443 & 2 & b  \\
40 & 12 & 55 & 27.92 & -11 & 52 & 54.3 & 0.1141 & 4 & 0.05 & 0.06 & 0.15E12 & 4.01 & 0.0136 & 0.0674 & 2 & b  \\
41 & 13 & 38 & 56.05 & -11 & 50 & 54.0 & 0.0521 & 3 & 0.03 & 0.04 & 0.61E11 & 2.55 & 0.0134 & 0.0335 & 2 & b  \\
42 & 13 & 59 & 03.99 & -11 & 24 & 02.2 & 0.0378 & 3 & 0.05 & 0.06 & 0.24E11 & 2.62 & 0.0303 & 0.0561 & 2 & b  \\
43 & 13 & 59 & 07.57 & -11 & 19 & 48.3 & 0.0385 & 4 & 0.06 & 0.04 & 0.38E11 & 2.62 & 0.0424 & 0.0801 & 2 & ab  \\
44 & 13 & 59 & 09.10 & -03 & 30 & 16.2 & 0.0777 & 3 & 0.04 & 0.04 & 0.57E11 & 1.66 & 0.0118 & 0.0420 & 2 & b  \\
45 & 13 & 59 & 39.23 & -12 & 00 & 37.5 & 0.0609 & 3 & 0.04 & 0.05 & 0.25E11 & 1.74 & 0.0146 & 0.0421 & 2 & b  \\
46 & 14 & 05 & 02.25 & -02 & 21 & 21.1 & 0.1398 & 3 & 0.05 & 0.05 & 0.22E12 & 3.84 & 0.0087 & 0.0506 & 2 & b  \\
47 & 14 & 24 & 18.60 & -03 & 16 & 20.8 & 0.0802 & 3 & 0.04 & 0.04 & 0.37E11 & 1.48 & 0.0100 & 0.0367 & 2 & b  \\
48 & 14 & 32 & 03.24 & -11 & 36 & 40.3 & 0.0597 & 3 & 0.05 & 0.06 & 0.40E11 & 1.58 & 0.0189 & 0.0533 & 2 & b  \\
49 & 14 & 34 & 34.73 & -03 & 33 & 09.2 & 0.0829 & 3 & 0.05 & 0.05 & 0.90E11 & 3.49 & 0.0118 & 0.0447 & 1 & b  \\
50 & 14 & 39 & 54.52 & -03 & 43 & 10.0 & 0.0594 & 4 & 0.04 & 0.04 & 0.56E11 & 1.88 & 0.0154 & 0.0434 & 2 & ab  \\
\noalign{\smallskip}
\hline
\end{tabular} }
\end{center}\end{flushleft}
\end{center}
\end{table*}

\addtocounter{table}{-1}
\newpage
\begin{table*}[ht]
\begin{center}
 \caption[c]{The main properties of LCCGs (continued)}
 \begin{flushleft}
 \begin{center} {\small
\begin{tabular}{lc@{\,}c@{\,}c@{ }c@{\,}c@{\,} ccccccccccc}
\hline
\noalign{\smallskip}
N & \multicolumn{3}{c}{RA} & \multicolumn{3}{c}{DEC} & $z_{cmb}$  & $N_{tot}$ & $R_{P}$ &$R_{h}$ & L & $L_{rat}$ & $\theta_{G}$ & $R_{G}$& Type & Notes \\
 \noalign{\smallskip}
&\multicolumn{6}{c}{(1950)}& & &\multicolumn{2}{c}{Mpc~$h^{-1}$}  & $L_{\odot}~h^{-2}$ & & &Mpc~$h^{-1}$ &&\\
 \noalign{\smallskip}
(1) & \multicolumn{3}{c}{(2)} & \multicolumn{3}{c}{(3)} & (4) & (5) & (6) & (7) & (8) & (9) & (10) & (11) & (l2) & (13)  \\
 \noalign{\smallskip}
\hline
 \noalign{\smallskip}
51 & 14 & 45 & 53.00 & -03 & 02 & 52.0 & 0.0721 & 4 & 0.05 & 0.05 & 0.38E11 & 1.43 & 0.0153 & 0.0512 & 2 & b  \\
52 & 14 & 50 & 56.82 & -02 & 42 & 21.6 & 0.0751 & 3 & 0.04 & 0.04 & 0.32E11 & 1.42 & 0.0114 & 0.0395 & 2 & b  \\
53 & 15 & 02 & 13.21 & -11 & 17 & 40.0 & 0.0812 & 3 & 0.03 & 0.04 & 0.50E11 & 1.66 & 0.0092 & 0.0339 & 2 & b  \\
54 & 15 & 12 & 28.56 & -02 & 35 & 25.1 & 0.1144 & 3 & 0.06 & 0.06 & 0.18E12 & 2.89 & 0.0110 & 0.0545 & 2 & b  \\
55 & 21 & 15 & 47.19 & -42 & 01 & 53.5 & 0.1086 & 3 & 0.07 & 0.07 & 0.71E11 & 2.38 & 0.0141 & 0.0668 & 2 & b  \\
56 & 21 & 18 & 31.69 & -45 & 42 & 08.5 & 0.0825 & 6 & 0.04 & 0.04 & 0.11E12 & 1.94 & 0.0143 & 0.0536 & 2 & b  \\
57 & 21 & 19 & 32.20 & -39 & 44 & 04.4 & 0.0589 & 3 & 0.06 & 0.07 & 0.50E11 & 1.96 & 0.0242 & 0.0674 & 2 & ab  \\
58 & 21 & 40 & 03.50 & -45 & 41 & 58.0 & 0.0564 & 3 & 0.04 & 0.05 & 0.19E11 & 2.54 & 0.0173 & 0.0465 & 2 & b  \\
59 & 21 & 49 & 26.48 & -42 & 09 & 48.6 & 0.0635 & 3 & 0.07 & 0.08 & 0.32E11 & 1.73 & 0.0270 & 0.0806 & 2 & b  \\
60 & 21 & 49 & 47.88 & -42 & 10 & 19.3 & 0.0627 & 3 & 0.04 & 0.04 & 0.26E11 & 1.76 & 0.0146 & 0.0431 & 2 & b  \\
61 & 21 & 57 & 06.40 & -44 & 29 & 56.7 & 0.1369 & 3 & 0.05 & 0.05 & 0.24E12 & 5.06 & 0.0087 & 0.0499 & 2 & b  \\
62 & 21 & 57 & 18.48 & -39 & 24 & 51.9 & 0.0598 & 3 & 0.05 & 0.05 & 0.31E11 & 2.10 & 0.0206 & 0.0583 & 2 & b  \\
63 & 22 & 03 & 03.62 & -38 & 20 & 38.4 & 0.1107 & 3 & 0.05 & 0.06 & 0.73E11 & 2.77 & 0.0113 & 0.0546 & 2 & b  \\
64 & 22 & 05 & 17.85 & -44 & 19 & 43.5 & 0.0738 & 3 & 0.03 & 0.04 & 0.61E11 & 2.42 & 0.0103 & 0.0352 & 2 & b  \\
65 & 22 & 08 & 10.33 & -45 & 41 & 43.2 & 0.0574 & 3 & 0.06 & 0.06 & 0.38E11 & 2.49 & 0.0225 & 0.0611 & 2 & ab  \\
66 & 22 & 32 & 51.14 & -41 & 45 & 22.2 & 0.0749 & 3 & 0.05 & 0.06 & 0.97E11 & 2.24 & 0.0134 & 0.0462 & 2 & b  \\
67 & 22 & 35 & 58.12 & -41 & 41 & 01.0 & 0.0935 & 4 & 0.05 & 0.01 & 0.91E11 & 2.77 & 0.0124 & 0.0520 & 2 & b  \\
68 & 22 & 42 & 37.18 & -45 & 19 & 55.3 & 0.0886 & 4 & 0.06 & 0.06 & 0.14E12 & 3.60 & 0.0163 & 0.0651 & 2 & b  \\
69 & 22 & 47 & 20.97 & -42 & 12 & 13.0 & 0.0862 & 3 & 0.05 & 0.05 & 0.60E11 & 2.33 & 0.0138 & 0.0540 & 2 & b  \\
70 & 22 & 47 & 45.66 & -45 & 37 & 41.4 & 0.0507 & 4 & 0.05 & 0.06 & 0.69E11 & 3.82 & 0.0263 & 0.0640 & 2 & b  \\
71 & 22 & 59 & 13.84 & -41 & 33 & 06.9 & 0.0451 & 3 & 0.06 & 0.06 & 0.31E11 & 2.40 & 0.0282 & 0.0617 & 2 & b  \\
72 & 23 & 16 & 37.14 & -42 & 15 & 28.2 & 0.0503 & 4 & 0.03 & 0.03 & 0.73E11 & 4.85 & 0.0124 & 0.0299 & 2 & b  \\
73 & 23 & 17 & 50.01 & -42 & 04 & 11.4 & 0.0562 & 3 & 0.03 & 0.03 & 0.13E12 & 4.44 & 0.0116 & 0.0310 & 2 & b  \\
74 & 23 & 33 & 22.26 & -38 & 51 & 26.5 & 0.0638 & 3 & 0.02 & 0.03 & 0.49E11 & 2.55 & 0.0080 & 0.0239 & 2 & b  \\
75 & 23 & 35 & 36.46 & -38 & 44 & 48.7 & 0.0550 & 3 & 0.06 & 0.07 & 0.55E11 & 2.73 & 0.0259 & 0.0678 & 2 & b  \\
76 & 23 & 59 & 44.84 & -44 & 16 & 49.3 & 0.0387 & 7 & 0.06 & 0.06 & 0.81E11 & 3.07 & 0.0375 & 0.0711 & 2 & ab  \\
\noalign{\smallskip}
\hline
\noalign{\smallskip}
(1) & \multicolumn{8}{l}{Running group ID number.}&(8) & \multicolumn{7}{l}{Luminosity in the LCRS $R$-band}\\
(2) & \multicolumn{8}{l}{Right ascension}&(9) & \multicolumn{7}{l}{Observed-to-total luminosity correction factor}\\
(3) & \multicolumn{8}{l}{Declination}&(10) & \multicolumn{7}{l}{Radius of the smallest circle including all members (in deg)}\\
(4) & \multicolumn{8}{l}{Redshift}&(11) & \multicolumn{7}{l}{Radius of the smallest circle including all members (in Mpc)}\\
(5) & \multicolumn{8}{l}{Number of group member galaxies}&(12) & \multicolumn{7}{l}{The type of CG (see text)}\\
(6) & \multicolumn{8}{l}{Mean pairwise separation}&(13) & \multicolumn{7}{l}{Comments}\\
(7) & \multicolumn{16}{l}{Harmonic radius} \\
\noalign{\smallskip}
\hline
\end{tabular} }
\end{center}\end{flushleft}
\end{center}
\end{table*}


%
\begin{table}{ }
 \caption[]{Compact Group Members}
 \label{tab:Cgmem}
 \begin{flushleft}
 \begin{center} {\small
\begin{tabular}{@{ }l@{ }c@{ }c@{ }c@{ }c@{ }c@{ }c@{ }c@{ }c@{ }c@{ }ccl@{ }c@{ }c@{ }c@{ }c@{ }c@{ }c@{ }c@{ }c@{ }c@{ }c@{ }}
  \noalign{\smallskip}
\cline{1-11} \cline{13-23}
\noalign{\smallskip}
ID&\multicolumn{3}{c}{RA}& \multicolumn{3}{c}{Dec}&mag
&\multicolumn{3}{c}{velocity} & &
ID&\multicolumn{3}{c}{RA}& \multicolumn{3}{c}{Dec}&mag
&\multicolumn{3}{c}{velocity}
\\ \cline{9-11} \cline{21-23}
& & &  && & &&vel &$z_{photo}$   &  Notes && & & &  && & &&vel &$z_{photo}$   &  Notes\\
\noalign{\smallskip}
(1)&\multicolumn{3}{c}{(2)}& \multicolumn{3}{c}{(3)}&(4)
&(5)&(6)&(7)&&(1)&\multicolumn{3}{c}{(2)}& \multicolumn{3}{c}{(3)}&(4)
&(5)&(6)&(7)\\
\noalign{\smallskip}
\cline{1-11} \cline{13-23}
\noalign{\smallskip}
01a& 00 & 22 & 03.48 & -45 & 34 & 24.2 & 15.57 & 20298 & 0.06 & m & & 15a& 04 & 16 & 59.59 & -42 & 16 & 35.1 & 15.84 & 16686 & 0.07 &c\\
01b& 00 & 22 & 06.11 & -45 & 34 & 32.8 & 15.90 & 20039 & 0.07 & c & & 15b& 04 & 17 & 03.90 & -42 & 17 & 26.9 & 16.83 & 16836 & 0.10 &c\\
01c& 00 & 22 & 04.22 & -45 & 33 & 50.3 & 17.18 & 20202 & 0.11 & m & & 15c& 04 & 17 & 01.45 & -42 & 17 & 33.6 & 17.32 & 16401 & 0.12 &m\\
02a& 00 & 45 & 51.32 & -42 & 16 & 55.0 & 16.40 & 23516 & 0.09 & m & & 16a& 04 & 17 & 07.56 & -42 & 19 & 13.2 & 15.01 & 15854 & 0.04 &c\\
02b& 00 & 45 & 49.74 & -42 & 16 & 20.7 & 16.85 & 23670 & 0.10 & c & & 16b& 04 & 17 & 05.22 & -42 & 18 & 49.8 & 15.44 & 15976 & 0.06 &m\\
02c& 00 & 45 & 48.09 & -42 & 17 & 19.6 & 16.86 & 23690 & 0.10 & c & & 16c& 04 & 17 & 12.78 & -42 & 19 & 44.8 & 17.30 & 16571 & 0.12 &c\\
02d& 00 & 45 & 49.05 & -42 & 17 & 48.1 & 17.41 & 23528 & 0.12 & m & & 17a& 04 & 19 & 34.35 & -45 & 22 & 03.7 & 17.07 & 41712 & 0.11 &b\\
03a& 01 & 04 & 08.69 & -44 & 35 & 11.2 & 17.01 & 41687 & 0.11 & m & & 17b& 04 & 19 & 35.42 & -45 & 22 & 12.1 & 17.21 & 41921 & 0.11 &m\\
03b& 01 & 04 & 08.09 & -44 & 35 & 21.6 & 17.31 & 41731 & 0.12 & b & & 17c& 04 & 19 & 33.92 & -45 & 22 & 14.1 & 17.51 & 41569 & 0.12 &m\\
03c& 01 & 04 & 06.95 & -44 & 35 & 02.5 & 17.66 & 41621 & 0.13 & m & & 18a& 10 & 04 & 30.99 & -05 & 22 & 45.8 & 15.18 & 17827 & 0.05 &m\\
04a& 01 & 21 & 22.10 & -39 & 30 & 04.0 & 16.93 & 36618 & 0.10 & c & & 18b& 10 & 04 & 29.17 & -05 & 22 & 53.5 & 16.52 & 17486 & 0.09 &c\\
04b& 01 & 21 & 20.31 & -39 & 29 & 42.7 & 17.34 & 36894 & 0.12 & m & & 18c& 10 & 04 & 30.15 & -05 & 22 & 30.4 & 17.19 & 17452 & 0.11 &m\\
04c& 01 & 21 & 18.95 & -39 & 29 & 47.9 & 17.63 & 36597 & 0.13 & m & & 19a& 10 & 04 & 51.61 & -05 & 26 & 30.0 & 15.84 & 25449 & 0.07 &m\\
05a& 01 & 39 & 08.14 & -44 & 50 & 57.9 & 16.80 & 28497 & 0.10 & c & & 19b& 10 & 04 & 54.19 & -05 & 26 & 48.5 & 15.85 & 25154 & 0.07 &b\\
05b& 01 & 39 & 05.07 & -44 & 50 & 56.1 & 17.15 & 28395 & 0.11 & m & & 19c& 10 & 04 & 53.07 & -05 & 26 & 29.2 & 17.62 & 25102 & 0.13 &m\\
05c& 01 & 39 & 08.81 & -44 & 51 & 13.0 & 17.51 & 28861 & 0.12 & m & & 20a& 10 & 17 & 28.57 & -03 & 37 & 47.2 & 16.05 & 20997 & 0.08 &m\\
06a& 02 & 00 & 25.45 & -45 & 01 & 31.2 & 15.99 & 18933 & 0.07 & m & & 20b& 10 & 17 & 27.59 & -03 & 37 & 32.6 & 16.77 & 21241 & 0.10 &e\\
06b& 02 & 00 & 25.95 & -45 & 01 & 01.7 & 17.48 & 18548 & 0.12 & m & & 20c& 10 & 17 & 29.99 & -03 & 37 & 21.1 & 16.94 & 21245 & 0.10 &m\\
06c& 02 & 00 & 22.90 & -45 & 01 & 24.4 & 17.52 & 18748 & 0.12 & c & & 21a& 10 & 20 & 42.59 & -05 & 37 & 21.1 & 16.29 & 26360 & 0.08 &c\\
07a& 02 & 27 & 02.11 & -44 & 40 & 16.4 & 16.60 & 29379 & 0.09 & m & & 21b& 10 & 20 & 42.12 & -05 & 37 & 36.2 & 16.69 & 26532 & 0.10 &m\\
07b& 02 & 26 & 59.87 & -44 & 41 & 01.5 & 16.63 & 29180 & 0.10 & c & & 21c& 10 & 20 & 43.71 & -05 & 37 & 12.9 & 17.00 & 26709 & 0.11 &m\\
07c& 02 & 27 & 02.54 & -44 & 40 & 48.9 & 17.07 & 29287 & 0.11 & m & & 22a& 10 & 26 & 22.02 & -02 & 57 & 56.3 & 15.89 & 11157 & 0.07 &c\\
08a& 02 & 50 & 10.57 & -39 & 06 & 27.8 & 16.35 & 31433 & 0.09 & m & & 22b& 10 & 26 & 20.61 & -02 & 59 & 24.3 & 16.88 & 10509 & 0.10 &b\\
08b& 02 & 50 & 12.01 & -39 & 06 & 49.9 & 16.44 & 31723 & 0.09 & b & & 22c& 10 & 26 & 21.34 & -02 & 59 & 45.3 & 16.91 & 10526 & 0.10 &m\\
08c& 02 & 50 & 13.10 & -39 & 06 & 43.7 & 17.44 & 31723 & 0.12 & m & & 23a& 10 & 27 & 34.21 & -02 & 55 & 06.7 & 15.17 & 10741 & 0.05 &c\\
09a& 02 & 53 & 16.75 & -39 & 27 & 05.7 & 15.18 & 19585 & 0.05 & m & & 23b& 10 & 27 & 39.72 & -02 & 54 & 00.1 & 16.93 & 11419 & 0.10 &c\\
09b& 02 & 53 & 19.25 & -39 & 26 & 06.0 & 15.64 & 19373 & 0.06 & m & & 23c& 10 & 27 & 39.91 & -02 & 54 & 55.3 & 17.15 & 11342 & 0.11 &c\\
09c& 02 & 53 & 16.34 & -39 & 26 & 28.8 & 17.54 & 19533 & 0.12 & c & & 24a& 10 & 43 & 30.45 & -02 & 19 & 00.3 & 17.00 & 32500 & 0.11 &c\\
10a& 03 & 18 & 31.98 & -42 & 26 & 55.5 & 16.92 & 34266 & 0.10 & m & & 24b& 10 & 43 & 31.75 & -02 & 18 & 28.7 & 17.17 & 32202 & 0.11 &m\\
10b& 03 & 18 & 31.94 & -42 & 26 & 09.9 & 17.37 & 34195 & 0.12 & m & & 24c& 10 & 43 & 30.06 & -02 & 18 & 40.9 & 17.61 & 32658 & 0.13 &m\\
10c& 03 & 18 & 33.25 & -42 & 26 & 36.6 & 17.62 & 34138 & 0.13 & c & & 25a& 10 & 44 & 07.17 & -11 & 22 & 24.4 & 16.08 & 16911 & 0.08 &c\\
11a& 03 & 20 & 30.80 & -41 & 31 & 30.1 & 17.11 & 18030 & 0.11 & m & & 25b& 10 & 44 & 03.95 & -11 & 21 & 53.7 & 16.56 & 15996 & 0.09 &c\\
11b& 03 & 20 & 23.54 & -41 & 32 & 16.9 & 17.58 & 18153 & 0.13 & m & & 25c& 10 & 44 & 04.90 & -11 & 21 & 43.2 & 17.03 & 15926 & 0.11 &m\\
11c& 03 & 20 & 28.66 & -41 & 32 & 06.2 & 17.76 & 18131 & 0.13 & m & & 26a& 11 & 09 & 49.08 & -02 & 53 & 17.9 & 15.59 & 24652 & 0.06 &c\\
11d& 03 & 20 & 28.03 & -41 & 32 & 13.3 & 17.80 & 18097 & 0.13 & c & & 26b& 11 & 09 & 48.04 & -02 & 53 & 20.4 & 16.85 & 25006 & 0.10 &m\\
12a& 03 & 30 & 26.94 & -38 & 18 & 23.7 & 15.65 & 17780 & 0.06 & m & & 26c& 11 & 09 & 47.34 & -02 & 52 & 51.7 & 16.99 & 24475 & 0.11 &m\\
12b& 03 & 30 & 26.64 & -38 & 18 & 04.8 & 16.12 & 17624 & 0.08 & c & & 27a& 11 & 12 & 15.49 & -06 & 05 & 33.4 & 16.15 & 24411 & 0.08 &m\\
12c& 03 & 30 & 26.15 & -38 & 17 & 51.8 & 16.93 & 17507 & 0.10 & m & & 27c& 11 & 12 & 12.52 & -06 & 06 & 00.1 & 16.24 & 24236 & 0.08 &m\\
13a& 03 & 33 & 25.61 & -38 & 49 & 53.2 & 15.68 & 18299 & 0.07 & m & & 27c& 11 & 12 & 14.70 & -06 & 05 & 14.6 & 16.61 & 23980 & 0.09 &m\\
13b& 03 & 33 & 22.19 & -38 & 49 & 21.4 & 16.30 & 18632 & 0.08 & c & & 27d& 11 & 12 & 14.50 & -06 & 05 & 52.2 & 16.22 & 24010 & 0.08 &c\\
13c& 03 & 33 & 20.68 & -38 & 51 & 04.7 & 16.99 & 18394 & 0.11 & m & & 28a& 11 & 12 & 48.85 & -03 & 38 & 16.2 & 16.38 & 24016 & 0.09 &m\\
13d& 03 & 33 & 22.15 & -38 & 50 & 21.1 & 17.50 & 18319 & 0.12 & b & & 28b& 11 & 12 & 49.80 & -03 & 38 & 21.7 & 17.06 & 24161 & 0.11 &m\\
14a& 03 & 57 & 52.19 & -41 & 27 & 33.9 & 15.05 & 17749 & 0.05 & c & & 28c& 11 & 12 & 50.76 & -03 & 37 & 10.6 & 17.14 & 24480 & 0.11 &c\\
14b& 03 & 57 & 43.66 & -41 & 28 & 18.1 & 15.80 & 17096 & 0.07 & m & & 28d& 11 & 12 & 50.65 & -03 & 37 & 41.6 & 17.31 & 24849 & 0.12 &m\\
14c& 03 & 57 & 48.40 & -41 & 28 & 10.4 & 16.04 & 17567 & 0.08 & c & & 28e& 11 & 12 & 48.61 & -03 & 38 & 53.3 & 17.43 & 24122 & 0.12 &c\\
\noalign{\smallskip}
\cline{1-11} \cline{13-23}
\end{tabular} }
\end{center}\end{flushleft}
\end{table}

\addtocounter{table}{-1}

\begin{table}[tbp]
 \caption [c]{Continued}
 \begin{flushleft}
 \begin{center} {\small
\begin{tabular}{@{ }l@{ }c@{ }c@{ }c@{ }c@{ }c@{ }c@{ }c@{ }c@{ }c@{ }ccl@{ }c@{ }c@{ }c@{ }c@{ }c@{ }c@{ }c@{ }c@{ }c@{ }c@{ }}
  \noalign{\smallskip}
\cline{1-11} \cline{13-23}
\noalign{\smallskip}
ID&\multicolumn{3}{c}{RA}& \multicolumn{3}{c}{Dec}&mag
&\multicolumn{3}{c}{velocity} & &
ID&\multicolumn{3}{c}{RA}& \multicolumn{3}{c}{Dec}&mag
&\multicolumn{3}{c}{velocity}
\\ \cline{9-11} \cline{21-23}
& & &  && & &&vel &$z_{photo}$   &  Notes && & & &  && & &&vel &$z_{photo}$   &  Notes\\
\noalign{\smallskip}
(1)&\multicolumn{3}{c}{(2)}& \multicolumn{3}{c}{(3)}&(4)
&(5)&(6)&(7)&&(1)&\multicolumn{3}{c}{(2)}& \multicolumn{3}{c}{(3)}&(4)
&(5)&(6)&(7)\\
\noalign{\smallskip}
\cline{1-11} \cline{13-23}
\noalign{\smallskip}
29a& 11 & 13 & 19.49 & -11 & 17 & 55.6 & 16.07 & 31626 & 0.08 & m & & 43a& 13 & 59 & 05.80 & -11 & 20 & 06.9 & 15.02 & 11137 & 0.04 &c\\
29b& 11 & 13 & 19.78 & -11 & 17 & 22.9 & 17.18 & 31646 & 0.11 & b & & 43b& 13 & 59 & 07.03 & -11 & 21 & 19.4 & 16.04 & 11243 & 0.08 &c\\
29c& 11 & 13 & 20.28 & -11 & 17 & 12.7 & 17.58 & 31735 & 0.13 & m & & 43c& 13 & 59 & 08.39 & -11 & 19 & 00.3 & 16.85 & 11267 & 0.10 &c\\
30a& 11 & 18 & 31.25 & -02 & 28 & 17.8 & 16.35 & 18710 & 0.09 & c & & 43d& 13 & 59 & 08.97 & -11 & 18 & 49.9 & 17.62 & 11475 & 0.13 &m\\
30b& 11 & 18 & 31.74 & -02 & 28 & 47.6 & 16.45 & 18825 & 0.09 & m & & 44a& 13 & 59 & 08.70 & -03 & 30 & 14.9 & 15.71 & 23093 & 0.07 &m\\
30c& 11 & 18 & 30.71 & -02 & 28 & 26.5 & 16.80 & 18546 & 0.10 & m & & 44b& 13 & 59 & 07.86 & -03 & 30 & 13.2 & 16.48 & 22942 & 0.09 &m\\
31a& 11 & 32 & 17.36 & -12 & 21 & 36.1 & 16.54 & 34885 & 0.09 & m & & 44c& 13 & 59 & 10.61 & -03 & 30 & 20.1 & 17.41 & 23037 & 0.12 &c\\
31b& 11 & 32 & 19.30 & -12 & 21 & 42.5 & 17.04 & 35091 & 0.11 & m & & 45a& 13 & 59 & 40.77 & -12 & 00 & 19.4 & 16.27 & 18175 & 0.08 &m\\
31c& 11 & 32 & 19.84 & -12 & 22 & 10.6 & 17.07 & 34772 & 0.11 & c & & 45b& 13 & 59 & 38.13 & -12 & 00 & 31.7 & 17.05 & 17673 & 0.11 &m\\
32a& 11 & 37 & 35.59 & -03 & 05 & 00.9 & 16.45 & 14637 & 0.09 & m & & 45c& 13 & 59 & 38.74 & -12 & 01 & 01.7 & 17.07 & 18101 & 0.11 &b\\
32b& 11 & 37 & 38.13 & -03 & 04 & 42.7 & 16.46 & 14709 & 0.09 & b & & 46a& 14 & 05 & 02.75 & -02 & 21 & 06.5 & 16.93 & 41583 & 0.10 &b\\
32c& 11 & 37 & 35.64 & -03 & 05 & 04.1 & 16.62 & 14557 & 0.09 & m & & 46b& 14 & 05 & 01.74 & -02 & 21 & 33.3 & 17.12 & 41551 & 0.11 &m\\
33a& 11 & 48 & 37.18 & -02 & 45 & 01.6 & 16.03 & 27412 & 0.08 & c & & 46c& 14 & 05 & 02.24 & -02 & 21 & 24.0 & 17.43 & 41821 & 0.12 &m\\
33b& 11 & 48 & 37.75 & -02 & 44 & 55.5 & 16.83 & 27166 & 0.10 & m & & 47a& 14 & 24 & 17.51 & -03 & 16 & 24.1 & 16.21 & 23672 & 0.08 &m\\
33c& 11 & 48 & 39.18 & -02 & 45 & 07.5 & 17.23 & 27388 & 0.11 & m & & 47b& 14 & 24 & 19.88 & -03 & 16 & 27.1 & 17.07 & 24068 & 0.11 &m\\
34a& 11 & 55 & 33.97 & -02 & 16 & 33.0 & 15.53 & 24410 & 0.06 & c & & 47c& 14 & 24 & 18.34 & -03 & 16 & 11.3 & 17.29 & 23695 & 0.12 &c \\
34b& 11 & 55 & 36.55 & -02 & 16 & 44.5 & 16.25 & 23776 & 0.09 & m & & 48a& 14 & 32 & 02.64 & -11 & 37 & 07.6 & 15.20 & 17666 & 0.05 &c\\
34c& 11 & 55 & 36.89 & -02 & 16 & 01.9 & 16.70 & 24522 & 0.10 & m & & 48b& 14 & 32 & 04.14 & -11 & 36 & 56.0 & 17.01 & 17427 & 0.11 &m\\
34d& 11 & 55 & 38.26 & -02 & 17 & 12.1 & 17.52 & 23803 & 0.12 & c & & 48c& 14 & 32 & 02.95 & -11 & 36 & 00.7 & 17.26 & 17882 & 0.12 &b\\
34e& 11 & 55 & 37.09 & -02 & 16 & 51.4 & 17.62 & 23846 & 0.13 & m & & 49a& 14 & 34 & 35.75 & -03 & 32 & 51.4 & 16.23 & 24460 & 0.08 &c\\
35a& 11 & 58 & 18.63 & -11 & 23 & 50.2 & 17.08 & 24863 & 0.11 & m & & 49b& 14 & 34 & 34.53 & -03 & 33 & 28.9 & 17.10 & 24539 & 0.11 &m\\
35b& 11 & 58 & 19.06 & -11 & 24 & 23.1 & 16.66 & 25131 & 0.10 & c & & 49c& 14 & 34 & 33.96 & -03 & 33 & 06.8 & 17.36 & 24895 & 0.12 &m\\
35c& 11 & 58 & 21.05 & -11 & 24 & 22.1 & 17.23 & 25298 & 0.11 & m & & 50a& 14 & 39 & 53.54 & -03 & 42 & 58.9 & 15.60 & 17406 & 0.06 &m\\
36a& 12 & 14 & 27.80 & -02 & 55 & 40.6 & 17.46 & 32019 & 0.12 & m & & 50b& 14 & 39 & 56.26 & -03 & 43 & 10.6 & 15.92 & 17663 & 0.07 &c\\
36b& 12 & 14 & 29.36 & -02 & 54 & 51.1 & 16.46 & 32100 & 0.09 & m & & 50c& 14 & 39 & 52.67 & -03 & 43 & 03.5 & 16.56 & 17326 & 0.09 &c\\
36c& 12 & 14 & 29.59 & -02 & 55 & 18.7 & 16.99 & 32183 & 0.11 & m & & 50d& 14 & 39 & 55.49 & -03 & 43 & 25.8 & 17.41 & 17908 & 0.12 &m\\
36d& 12 & 14 & 29.77 & -02 & 55 & 28.4 & 17.59 & 32171 & 0.13 & c & & 51a& 14 & 45 & 51.64 & -03 & 02 & 25.7 & 16.48 & 21510 & 0.09 &m\\
36e& 12 & 14 & 31.09 & -02 & 55 & 22.7 & 17.00 & 32154 & 0.11 & m & & 51b& 14 & 45 & 52.70 & -03 & 03 & 14.2 & 16.49 & 21202 & 0.09 &m\\
37a& 12 & 40 & 36.11 & -11 & 57 & 34.2 & 16.34 & 28878 & 0.09 & c & & 51c& 14 & 45 & 54.51 & -03 & 02 & 52.5 & 16.93 & 21399 & 0.10 &m\\
37b& 12 & 40 & 35.08 & -11 & 57 & 16.8 & 16.83 & 28684 & 0.10 & m & & 51d& 14 & 45 & 53.00 & -03 & 02 & 56.0 & 17.23 & 21427 & 0.11 &c\\
37c& 12 & 40 & 34.79 & -11 & 57 & 27.9 & 17.01 & 28789 & 0.11 & m & & 52a& 14 & 50 & 58.04 & -02 & 42 & 11.1 & 16.00 & 22262 & 0.08 &m\\
38a& 12 & 47 & 00.17 & -06 & 04 & 43.2 & 16.15 & 37525 & 0.08 & m & & 52b& 14 & 50 & 55.64 & -02 & 42 & 29.4 & 17.09 & 22441 & 0.11 &c\\
38b& 12 & 47 & 00.88 & -06 & 04 & 16.8 & 17.29 & 37710 & 0.12 & c & & 52c& 14 & 50 & 56.80 & -02 & 42 & 24.0 & 17.68 & 22174 & 0.13 &m\\
38c& 12 & 47 & 00.08 & -06 & 04 & 58.7 & 17.34 & 36970 & 0.12 & m & & 53a& 15 & 02 & 13.47 & -11 & 17 & 35.4 & 15.87 & 24269 & 0.07 &m\\
39a& 12 & 50 & 24.92 & -11 & 48 & 37.6 & 17.22 & 30575 & 0.11 & c & & 53b& 15 & 02 & 12.77 & -11 & 17 & 26.5 & 17.21 & 24068 & 0.11 &m\\
39b& 12 & 50 & 23.55 & -11 & 48 & 42.5 & 17.31 & 30500 & 0.12 & m & & 53c& 15 & 02 & 13.41 & -11 & 17 & 57.9 & 17.34 & 24064 & 0.12 &c\\
39c& 12 & 50 & 22.76 & -11 & 48 & 51.9 & 17.58 & 30368 & 0.13 & m & & 54a& 15 & 12 & 28.76 & -02 & 35 & 16.2 & 15.85 & 34247 & 0.07 &c\\
40a& 12 & 55 & 27.61 & -11 & 53 & 02.2 & 17.28 & 34141 & 0.12 & m & & 54b& 15 & 12 & 27.66 & -02 & 35 & 13.7 & 17.36 & 33904 & 0.12 &m\\
40b& 12 & 55 & 29.41 & -11 & 53 & 04.0 & 17.35 & 33629 & 0.12 & m & & 54c& 15 & 12 & 29.24 & -02 & 35 & 44.8 & 17.36 & 34189 & 0.12 &m\\
40c& 12 & 55 & 28.09 & -11 & 52 & 51.7 & 17.59 & 33942 & 0.13 & m & & 55a& 21 & 15 & 48.63 & -42 & 02 & 13.6 & 17.18 & 32756 & 0.11 &m\\
40d& 12 & 55 & 26.57 & -11 & 52 & 39.2 & 17.67 & 33835 & 0.13 & b & & 55b& 21 & 15 & 47.67 & -42 & 01 & 48.0 & 17.19 & 32867 & 0.11 &b\\
41a& 13 & 38 & 55.33 & -11 & 50 & 32.4 & 15.34 & 15201 & 0.05 & m & & 55c& 21 & 15 & 45.31 & -42 & 01 & 39.2 & 17.40 & 32672 & 0.12 &m\\
41b& 13 & 38 & 56.05 & -11 & 51 & 19.1 & 15.64 & 15300 & 0.06 & m & & 56a& 21 & 18 & 29.42 & -45 & 42 & 02.8 & 15.60 & 24415 & 0.06 &m\\
41c& 13 & 38 & 56.76 & -11 & 50 & 48.3 & 17.08 & 15453 & 0.11 & c & & 56b& 21 & 18 & 31.86 & -45 & 41 & 46.6 & 16.59 & 25101 & 0.09 &m\\
42a& 13 & 59 & 01.37 & -11 & 23 & 18.9 & 15.59 & 11506 & 0.06 & c & & 56c& 21 & 18 & 32.20 & -45 & 42 & 36.2 & 17.12 & 24805 & 0.11 &m\\
42b& 13 & 59 & 03.76 & -11 & 24 & 16.0 & 16.27 & 10744 & 0.08 & m & & 56d& 21 & 18 & 31.18 & -45 & 41 & 54.5 & 17.43 & 25278 & 0.12 &m\\
42c& 13 & 59 & 06.83 & -11 & 24 & 31.9 & 17.13 & 10911 & 0.11 & c & & 56e& 21 & 18 & 32.64 & -45 & 42 & 25.8 & 17.69 & 24827 & 0.13 &c\\
\noalign{\smallskip}
\cline{1-11} \cline{13-23}
\end{tabular} }
\end{center}
\end{flushleft}
\end{table}

\addtocounter{table}{-1}
\begin{table}[tbp]
 \caption [c]{Continued}
 \begin{flushleft}
 \begin{center} {\small 
\begin{tabular}{@{ }l@{ }c@{ }c@{ }c@{ }c@{ }c@{ }c@{ }c@{ }c@{ }c@{ }ccl@{ }c@{ }c@{ }c@{ }c@{ }c@{ }c@{ }c@{ }c@{ }c@{ }c@{ }}
  \noalign{\smallskip}
\cline{1-11} \cline{13-23}
\noalign{\smallskip}
ID&\multicolumn{3}{c}{RA}& \multicolumn{3}{c}{Dec}&mag
&\multicolumn{3}{c}{velocity} & &
ID&\multicolumn{3}{c}{RA}& \multicolumn{3}{c}{Dec}&mag
&\multicolumn{3}{c}{velocity}
\\ 
\cline{9-11} \cline{21-23}
& & &  && & &&vel &$z_{photo}$   &  Notes && & & &  && & &&vel &$z_{photo}$   &  Notes\\
\noalign{\smallskip}
(1)&\multicolumn{3}{c}{(2)}& \multicolumn{3}{c}{(3)}&(4)
&(5)&(6)&(7)&&(1)&\multicolumn{3}{c}{(2)}& \multicolumn{3}{c}{(3)}&(4)
&(5)&(6)&(7)\\
\noalign{\smallskip}
\cline{1-11} \cline{13-23}
\noalign{\smallskip}
56f& 21 & 18 & 32.68 & -45 & 42 & 05.0 & 17.30 & 25189 & 0.12 & m & & 68a& 22 & 42 & 37.40 & -45 & 20 & 29.4 & 16.13 & 26456 & 0.08 &c\\
57a& 21 & 19 & 35.50 & -39 & 43 & 51.7 & 15.11 & 17949 & 0.05 & c & & 68b& 22 & 42 & 35.90 & -45 & 19 & 39.9 & 17.02 & 27258 & 0.11 &m\\ 
57b& 21 & 19 & 28.20 & -39 & 44 & 12.1 & 17.02 & 17945 & 0.11 & c & & 68c& 22 & 42 & 36.91 & -45 & 19 & 32.1 & 17.27 & 27142 & 0.12 &b\\ 
57c& 21 & 19 & 32.80 & -39 & 44 & 09.7 & 17.56 & 17754 & 0.12 & m & & 68d& 22 & 42 & 38.57 & -45 & 20 & 01.7 & 17.28 & 26310 & 0.12 &m\\
58a& 21 & 40 & 01.77 & -45 & 41 & 30.3 & 16.99 & 17126 & 0.11 & b & & 69a& 22 & 47 & 20.78 & -42 & 12 & 13.2 & 16.24 & 26066 & 0.08 &c\\ 
58b& 21 & 40 & 03.28 & -45 & 42 & 04.0 & 17.25 & 16997 & 0.11 & m & & 69b& 22 & 47 & 18.92 & -42 & 12 & 18.3 & 17.25 & 26112 & 0.11 &m\\
58c& 21 & 40 & 05.42 & -45 & 42 & 18.9 & 17.57 & 17185 & 0.12 & e & & 69c& 22 & 47 & 23.25 & -42 & 12 & 07.3 & 17.69 & 26080 & 0.13 &m\\ 
59a& 21 & 49 & 24.19 & -42 & 09 & 06.1 & 15.79 & 19209 & 0.07 & m & & 70a& 22 & 47 & 48.16 & -45 & 36 & 56.0 & 15.89 & 14997 & 0.07 &c\\
59b& 21 & 49 & 28.94 & -42 & 10 & 27.5 & 17.15 & 19573 & 0.11 & m & & 70b& 22 & 47 & 41.97 & -45 & 38 & 04.6 & 16.05 & 15564 & 0.08 &c\\ 
59c& 21 & 49 & 26.06 & -42 & 09 & 48.2 & 17.41 & 19042 & 0.12 & c & & 70c& 22 & 47 & 45.18 & -45 & 38 & 13.8 & 16.60 & 15328 & 0.10 &m\\
60a& 21 & 49 & 50.48 & -42 & 10 & 25.0 & 16.55 & 19122 & 0.09 & b & & 70d& 22 & 47 & 46.53 & -45 & 37 & 45.0 & 17.57 & 15752 & 0.12 &m\\
60b& 21 & 49 & 47.44 & -42 & 10 & 22.2 & 16.72 & 18965 & 0.10 & m & & 71a& 22 & 59 & 17.92 & -41 & 33 & 04.7 & 15.74 & 13656 & 0.07 &m\\
60c& 21 & 49 & 45.95 & -42 & 10 & 11.5 & 17.11 & 18982 & 0.11 & m & & 71b& 22 & 59 & 14.75 & -41 & 32 & 50.8 & 15.94 & 13812 & 0.07 &b\\ 
61a& 21 & 57 & 07.77 & -44 & 29 & 55.0 & 17.07 & 41262 & 0.11 & m & & 71c& 22 & 59 & 09.07 & -41 & 33 & 25.3 & 17.29 & 13856 & 0.12 &b\\ 
61b& 21 & 57 & 06.65 & -44 & 29 & 57.4 & 17.33 & 41173 & 0.12 & c & & 72a& 23 & 16 & 38.27 & -42 & 15 & 13.3 & 15.54 & 15420 & 0.06 &c\\ 
61c& 21 & 57 & 04.86 & -44 & 29 & 57.6 & 17.52 & 41332 & 0.12 & m & & 72b& 23 & 16 & 38.51 & -42 & 15 & 32.7 & 17.31 & 15600 & 0.12 &m\\
62a& 21 & 57 & 18.12 & -39 & 25 & 23.7 & 16.42 & 18309 & 0.09 & c & & 72c& 23 & 16 & 36.09 & -42 & 15 & 17.7 & 17.32 & 15152 & 0.12 &m\\
62b& 21 & 57 & 18.65 & -39 & 24 & 10.6 & 16.70 & 18172 & 0.10 & b & & 72d& 23 & 16 & 35.77 & -42 & 15 & 47.4 & 17.40 & 15154 & 0.12 &m\\
62c& 21 & 57 & 18.69 & -39 & 24 & 59.7 & 16.96 & 18055 & 0.11 & m & & 73a& 23 & 17 & 51.20 & -42 & 04 & 28.8 & 15.44 & 16989 & 0.06 &m\\
63a& 22 & 03 & 03.83 & -38 & 20 & 38.3 & 17.30 & 33284 & 0.12 & c & & 73b& 23 & 17 & 49.36 & -42 & 04 & 14.1 & 15.84 & 17193 & 0.07 &m\\ 
63b& 22 & 03 & 04.58 & -38 & 20 & 54.2 & 17.49 & 33719 & 0.12 & m & & 73c& 23 & 17 & 49.51 & -42 & 03 & 51.8 & 16.23 & 17103 & 0.08 &c\\
63c& 22 & 03 & 02.44 & -38 & 20 & 22.2 & 17.50 & 33303 & 0.12 & m & & 74a& 23 & 33 & 22.41 & -38 & 51 & 32.3 & 16.13 & 19484 & 0.08 &m\\ 
64a& 22 & 05 & 18.01 & -44 & 19 & 49.6 & 15.85 & 22193 & 0.07 & m & & 74b& 23 & 33 & 23.39 & -38 & 51 & 26.3 & 16.36 & 19470 & 0.09 &c\\
64b& 22 & 05 & 16.11 & -44 & 19 & 39.3 & 16.90 & 22510 & 0.10 & m & & 74c& 23 & 33 & 21.00 & -38 & 51 & 21.2 & 17.46 & 19156 & 0.12 &m\\ 
64c& 22 & 05 & 19.46 & -44 & 19 & 41.7 & 17.59 & 22327 & 0.13 & c & & 75a& 23 & 35 & 37.06 & -38 & 45 & 00.8 & 15.50 & 16560 & 0.06 &m\\ 
65a& 22 & 08 & 07.18 & -45 & 41 & 15.1 & 16.13 & 17521 & 0.08 & b & & 75b& 23 & 35 & 32.17 & -38 & 44 & 43.5 & 16.15 & 16928 & 0.08 &b\\
65b& 22 & 08 & 10.97 & -45 & 41 & 42.0 & 16.49 & 17371 & 0.09 & m & & 75c& 23 & 35 & 40.02 & -38 & 44 & 42.2 & 17.60 & 16747 & 0.13 &b\\
65c& 22 & 08 & 12.72 & -45 & 42 & 10.9 & 17.16 & 17361 & 0.11 & c & & 76a& 23 & 59 & 48.45 & -44 & 16 & 23.8 & 15.51 & 11935 & 0.06 &c\\ 
66a& 22 & 32 & 49.87 & -41 & 45 & 41.3 & 15.48 & 22652 & 0.06 & m & & 76b& 23 & 59 & 45.15 & -44 & 16 & 00.7 & 15.54 & 11916 & 0.06 &m\\
66b& 22 & 32 & 53.52 & -41 & 45 & 18.0 & 16.28 & 22605 & 0.08 & c & & 76c& 23 & 59 & 49.00 & -44 & 17 & 03.8 & 15.56 & 11563 & 0.06 &m\\ 
66c& 22 & 32 & 50.09 & -41 & 45 & 08.6 & 16.49 & 22807 & 0.09 & m & & 76d& 23 & 59 & 40.74 & -44 & 16 & 06.5 & 16.11 & 12356 & 0.08 &c\\ 
67a& 22 & 35 & 57.07 & -41 & 40 & 39.6 & 16.99 & 28506 & 0.11 & m & & 76e& 23 & 59 & 48.85 & -44 & 17 & 35.1 & 16.36 & 11415 & 0.09 &c\\
67b& 22 & 35 & 58.24 & -41 & 41 & 22.0 & 17.04 & 27996 & 0.11 & m & & 76f& 23 & 59 & 37.93 & -44 & 17 & 35.9 & 16.63 & 12203 & 0.10 &c\\
67c& 22 & 35 & 58.27 & -41 & 41 & 21.7 & 17.09 & 28332 & 0.11 & c & & 76g& 23 & 59 & 44.08 & -44 & 16 & 57.5 & 16.91 & 11485 & 0.10 &c\\ 
67d& 22 & 35 & 58.94 & -41 & 40 & 44.0 & 17.37 & 28199 & 0.12 & m & & 
  &    &    &       &     &    &      &       &       &      & \\
\noalign{\smallskip}
\cline{1-11} \cline{13-23}
\noalign{\smallskip}
(1) & \multicolumn{11}{l}{Compact group member}\\
(2) & \multicolumn{11}{l}{Right ascension}\\
(3) & \multicolumn{11}{l}{Declination}\\
(4) & \multicolumn{11}{l}{$R$-band magnitude}\\
(5) & \multicolumn{11}{l}{Velocity}\\
(6) & \multicolumn{11}{l}{$z_{\rm photo}$}\\
(7) &\multicolumn{11}{l}{Notes}\\
\noalign{\smallskip}
\cline{1-11} 
\end{tabular} }
\end{center}\end{flushleft}
\end{table}

\acknowledgements
This research has made use of the NASA/IPAC Extragalactic Database
(NED), which is operated by the Jet Propulsion Laboratory, at the
California Institute of Technology, under contract with the National
Aeronautics and Space Administration. The Digitized Sky Survey was 
produced at the Space Telescope Science Institute under US government
grant NAGW-2166.
\\
This work was supported in part by the US Department of Energy
under contract No.\ DE-AC02-76CH03000.

\refer
\aba
\rf{Allam, S., Tucker, D., Lin,H., Hashimoto, Y.: 1999, Astrophys. J. 522, L89}
\rf{Barton, E., Geller, M., Ramella, M., Marzke, R., de Costa, L.: 1996, Astron. J. 112, 871}
\rf{Diaferio, A., Geller, M.J., Ramella, M.: 1994, Astron. J. 107, 868}
\rf{Hernquist, L., Katz, N., Weinberg, D.H.: 1995, Astrophys. J., 442, 57}
\rf{Hickson, P.: 1982, Astrophys. J., 225, 382}
\rf{Mamon, G.A.: 1986, Astrophys. J., 307, 426}
\rf{Ramella, M., Geller, M.J., Huchra, J.P.: 1989, Astrophys. J., 344, 57}
\rf{Shectman, S., Landy, S., Oemler A., Tucker, D., Lin, H., Kirshner, R., Schechter, P.: 1996, Astrophys. J., 470, 172}
\rf{Tucker, D., Oemler, A., Hashimoto, Y., Shectman, S., Kirshner, R., Lin, H., Landy,S., Schechter, P., Allam, S.: 2000, Astrophys. J. Suppl., submitted}
\abe

\addresses 
\rf{Allam, S., Helwan, Cairo, Egypt, sallam@fnal.gov}
\rf{Tucker, D., Batavia, Illinois, USA, dtucker@fnal.gov}

\end{document}